\documentstyle[draft,epsf]{mn}
\def\etal{{\rm et al.\ }}
\def\eg{{\rm eg.\ }}
\def\ie{{\rm ie.\ }}
\def\cf{{\rm cf.\ }}
\def\vs{{\rm vs.\ }}
\def\hi{{\rm H{\sc i} }}
\def\oii{{\rm O{\sc ii} }}
\def\oiii{{\rm O{\sc iii} }}
\def\gsim{\mathrel{\raise0.35ex\hbox{$\scriptstyle >$}\kern-0.6em 
\lower0.40ex\hbox{{$\scriptstyle \sim$}}}}
\def\lsim{\mathrel{\raise0.35ex\hbox{$\scriptstyle <$}\kern-0.6em 
\lower0.40ex\hbox{{$\scriptstyle \sim$}}}}

\title{A Bayesian Classifier for Photometric Redshifts: \\
Identification of high redshift clusters}

\author[Kodama, Bell \& Bower]
{
Tadayuki Kodama$^1$, Eric F. Bell$^2$ \& Richard G. Bower$^2$\\
$^1$ Institute of Astronomy, University of Cambridge, Madingley Road, Cambridge CB3 0HA, UK\\
$^2$ Department of Physics, University of Durham, South Road, Durham DH1 3LE, UK
}

\begin{document}
\date{5 October 1998}

\label{firstpage}

\maketitle

\begin{abstract}
Photometric redshift classifiers provide a means of estimating galaxy
redshifts from observations using a small number of broad-band filters. 
However, the accuracy with which redshifts can be determined is sensitive
to the star formation history of the galaxy, for example the effects 
of age, metallicity and on-going star formation.
We present a photometric classifier that explicitly takes into account the
degeneracies implied by these variations, based on the flexible stellar
population synthesis code of Kodama \& Arimoto.
The situation is encouraging since many of the variations in stellar
populations introduce colour changes that are degenerate. We use a Bayesian
inversion scheme to estimate the likely range of redshifts compatible with
the observed colours.  When applied to existing multi-band photometry
for Abell~370, most of the cluster members are correctly recovered
with little field contamination.
The inverter is focussed on the recovery of a wide variety of galaxy
populations in distant ($z \sim 1$) clusters from broad band colours
covering the 4000~{\AA} break.  It is found that this can be achieved 
with impressive accuracy ($|\Delta z| < 0.1$), allowing detailed
investigation into the evolution of cluster galaxies with little selection
bias.
\end{abstract}


\section{Introduction}

The current trend in cosmology is to explore the properties of galaxies
at ever fainter limits. This has lead to demonstration of the existence
of a substantial galaxy population at $z>3$ (Steidel et al.\ 1996,
Metcalfe et al.\ 1996,  Steidel et al.\ 1998), and the 
discovery of galaxy clusters with $z\gsim1$ (Deltorn et al.\ 1997,
Stanford et al.\ 1997, Yamada et al.\ 1997).
These discoveries have allowed us to extend our 
knowledge of the formation history of galaxies (Madau et al.\ 1996, 
Baugh et al.\ 1998, Kodama et al.\ 1998) and the growth of the universe's 
gravitational structure (Bower \& Smail 1997). However, while images 
that reach these depths are now relatively 
commonplace, spectroscopic follow-up of these objects is extremely time 
consuming even on 8m-class telescopes. These problems are offset by the 
multiplex capability of multi-object spectrographs (eg. LDSS; Allington-Smith 
et al.\ 1994) and fibre-optic fed spectrographs (eg. Taylor 1995), or 
by surveys targeted at specific redshifts using tuneable narrow-band filters 
(eg. Jones \& Bland-Hawthorn 1997). Nevertheless, even in the best studied 
deep images, only a small fraction of the galaxies have known
spectroscopic redshifts.

Whereas spectroscopic redshifts use sharp absorption and/or emission lines 
to accurately determine the rest wavelength of the spectrum, it is 
also possible to exploit the overall characteristic shape of the 
spectral energy distribution (SED) to estimate the galaxy's redshift. 
This `photometric redshift' approach can be applied to broad band images
provided they have sufficiently high signal to noise and adequately sample
the important features of the SED. In particular, the 
4000~{\AA} spectral break and the Balmer and Lyman series limits 
are important features that
arise in almost all galaxy spectra. Although precise redshifts cannot be
determined by this method, estimates of (or limits on) $z$ are obtained.

The existing photometric redshift estimators fall into three main
classes: empirical redshift estimators, those based on observed spectral
energy distributions and model-based redshift estimators.  Empirical redshift 
estimators (Connolly et al.\ 1995) are based on a training set of 
galaxies for which the redshifts and broad-band fluxes are known. 
These are used to train an estimator, for example a multi-dimensional 
polynomial fit, that predicts the redshift from the input fluxes with 
minimum error. 
The disadvantage of this method is that it requires a relatively large 
training dataset with high quality colours and known redshifts. This makes
it difficult to apply beyond the limits of spectroscopic surveys, although
this problem might be alleviated using the colours of distant, gravitationally
lensed galaxies. 
However this method, when tested against independent but similar data, 
can give impressive accuracy ($\sigma_z$$\sim$0.06; Connolly et al.\ 1995).  

Lanzetta et al.\ (1996), Mobasher et al.\ (1996) and 
Sawicki et al.\ (1997) use an approach that is 
based on the observed SEDs of galaxies covering a wide range of spectral 
types. Redshifts
are estimated from the observed data by redshifting each of the templates
and determining the best match to the observational colours. They
emphasise the importance of using observed templates in order to
incorporate the effects of dust. This is particularly important for 
galaxies in the redshift range $1 < z < 3$ because the
optical colours increasingly reflect the rest frame ultraviolet spectrum
of the galaxy. One problem with this approach, however, is that the 
spectral library does not take into account the evolution of the galaxy
stellar populations. The method can accommodate evolution in as far as it 
is equivalent to changing galaxies between different spectral types, however.

Model-based approaches 
use stellar population synthesis codes
(\eg Bruzual \& Charlot 1993) to produce model SEDs that can then be 
compared to the observed data. 
For example, Gwyn \& Hartwick (1996) used a spread of galaxy models
from single burst stellar populations to models with constant star
formation to model present-day galaxies.
When generating redshifted model SEDs, the evolution 
of the stellar population is automatically taken into account.  The redshift
of the observed galaxy is determined by minimising $\chi^2$ residuals.  
The improved flexibility of this method can however, lead to greater
errors in the estimated redshifts. 
This arises because of colour degenaracies between the
effects of galaxy type and redshift.

In this paper, we focus more closely on the interrelation between 
star formation history and redshift estimation. As we have outlined,
photometric redshifts can be susceptible to changes in the
galaxy stellar population. For instance, the effects of 
age, metal abundance and
on-going star formation are all reflected in the relative shape 
of the continuum, particularly when it is convolved with the response
of standard broad band filters. It is important that these uncertainties
are taken into account when determining the galaxy redshift. 
We develop a method of photometric redshift determination that explicitly 
takes into account the degeneracies implied by these variations. 
Clearly, incorporating additional free parameters to describe the star
formation history of the galaxy threatens to make it impossible to
extract useful redshift information.  However, many of the changes in colour
caused by different star formation histories are degenerate:
this is the familiar age-metallicity degeneracy that has long plagued
the estimation of star formation histories in elliptical galaxies.
We will show that 
for red galaxies, redshifts can be determined under only weak assumptions 
about the star formation history. 
At lower redshifts, the colours of blue (disk)
galaxies become considerably harder to disentangle.

Our approach attempts to deal with, and indeed embrace, this unavoidable
degeneracy in colours with variations in redshift and star formation history.
We explicitly account for galaxy metallicities and star formation
histories; these effects are in many cases degenerate with 
uncertainties due to the stellar initial mass function (IMF), recent star
formation, dust extinction and cosmology.
We retain possible degeneracies in plausible values of
galaxy type and redshift by storing a `probability map' 
for each galaxy, which can
be used to estimate a range of acceptable redshifts rather than reducing 
the observed data to a single `best bet' estimate of galaxy
type and redshift. In particular, our classifier is designed to 
pick out galaxy cluster members without biasing the sample to galaxies 
of one particular star formation history. Our motivation is to use this 
method to study the {\it photometric} properties of $z\sim1$ cluster galaxies 
with as little selection bias as possible.

The structure of the paper is as follows. \S~2 introduces the stellar 
population synthesis code of Kodama \& Arimoto (1997). We derive colour 
tracks for a range of galaxy star formation histories and outline the major 
uncertainties in these tracks. This provides the framework for selecting
appropriate filter sets and required photometric accuracy.
\S~3 details our Bayesian approach to the inversion problem. We explicitly
incorporate a wide variety of possible star formation histories, and
explicitly incorporate the resulting degeneracies in our redshift estimates. 
The role of the prior is discussed.
In \S~4, we test our method with galaxies in Abell~370 cluster field and
galaxies with known redshifts in the Hubble Deep Field (HDF).
\S~5 gives an application of the method to a simulated cluster at $z=1$. 
A summary and our conclusions are presented in \S~6.

\section{Colour tracks as a function of star formation history}

\subsection{Model}

The evolutionary population synthesis model of Kodama (1997)
was used to predict the photometric properties
of evolving stellar populations.  This model calculates
the spectral evolution of a galaxy with an arbitrary star formation
history, taking into account the chemical evolution in a self-consistent
way.  Kodama \& Arimoto (1997) applied this model
to the elliptical galaxy populations of distant clusters.  In this
study, disk models with ongoing star formation are considered in 
addition to the elliptical models.  We first describe the basic 
equations and parameters of this model and then summarise the 
elliptical galaxy and disk galaxy models.

\subsubsection{Equations and parameters}

 We assume that the galactic gas is supplied from a
 surrounding gas reservoir trapped in the gravitational
 potential of a galaxy and that the gas is always well-mixed and
 distributes uniformly in a galaxy.
 The star formation is described by the following equations.
 The stellar IMF is given by a single
 power law:
 \begin{equation}
 \phi(m) = A m^{-x},\hspace{0.5cm} m_l \le m \le m_u,
 \end{equation}
 where $m_l$ and $m_u$ are lower and upper limits of
 initial stellar mass respectively. The Salpeter (1955)
 IMF corresponds to $x = 1.35$.
 The coefficient $A$ is determined by,
 \begin{equation}
 \int_{m_l}^{m_u} \phi(m)dm = 1.
 \end{equation}
 The IMF is assumed to be time invariant.
 The star formation rate (SFR) $\psi(t)$ is assumed to be proportional
 to the gas mass $M_g(t)$ (Schmidt 1959):
 \begin{equation}
 \psi(t) = \frac{1}{\tau} M_g(t),
 \end{equation}
 where $\tau$ is the star formation time scale in Gyr.
 Note that this formulation gives an exponentially decaying SFR with
 an effective time scale $\tau/\alpha$ in the case of the simple models,
 where $\alpha$ is the so-called locked-up mass fraction defined by Tinsley
 (1980). The Salpeter mass function ($x=1.35$) with $m_l=0.1$,
 $m_u=60$ M$_{\odot}$ gives $\alpha=0.72$.
 The gas infall rate $\xi_{in}(t)$ depends on the
 initial total mass of the gas reservoir $M_T$ and the gas infall time scale
 $\tau_{in}$:
 \begin{equation}
 \xi_{in}(t) =  \frac{M_T}{\tau_{in}} \exp(-\frac{t}{\tau_{in}})
 \end{equation}
 (\cf K\"oppen \& Arimoto 1990).
 The gas metallicity $Z_g(t)$ is calculated numerically, using the
 basic equations of chemical evolution (Tinsley 1980) and stellar
 nucleosynthesis tables (Nomoto 1993).
 The metal contribution from SNIa is also considered
 by fixing their lifetime at 1.5 Gyr.
 We assume that the 
 metal-enriched gas spreads through the galaxy instantaneously 
 and evenly (the one-zone approximation).
 As the initial conditions, we assume that there is no gas in a
 galaxy before the onset of star formation;
 \ie $M_g(0)=0$ and $Z_g(0)=0$.
 Using the infall history defined as above, our expression for the star
 formation rate $\psi(t)$ and the metallicity of the stars $Z(t) =
 Z_g(t)$, the integrated spectrum of a galaxy can be synthesised as a
 function of time.
 By specifying the galaxy age, or equivalently its formation redshift, 
 and cosmological parameters, we obtain the spectra and therefore
 colour indices of the galaxy as a function of redshift.
 The cosmological parameters are set to $H_0=50$ km s$^{-1}$ Mpc$^{-1}$,
 $\Omega_0=1.0$, and $\Lambda_0=0.0$ unless otherwise stated.

\subsubsection{Elliptical galaxies and bulges}
 For elliptical galaxy models (E models), we use $x=1.10$, 
 $m_{l}=0.1$ M$_{\odot}$, and $m_{u}=60$ M$_{\odot}$
 for the IMF and short time scales of star formation and gas
 infall: $\tau=\tau_{in}=0.1$ Gyr.
 The slope of the IMF differs
 from the Salpeter value $x=1.35$ to 
 allow the colours of the reddest giant ellipticals to be reproduced 
 in the context of this model (with a Salpeter IMF, metallicities
 high enough cannot be achieved).
 In addition, in order to reproduce the observed present-day
 dependence of elliptical galaxy colour on luminosity, it is
 useful to introduce another parameter, the 
 galactic wind epoch $t_{gw}$.
 At this time, the energy put into the ISM in the proto-elliptical
 galaxy by SNe is large enough to overcome the potential of the
 galaxy, resulting in the ejection of the gas from the galaxy, ending
 star formation.
 We constructed a model sequence of elliptical galaxies as a function of
 total luminosity by simply changing $t_{gw}$ so that they reproduce the
 colour-magnitude ({\it C-M}) relation of Coma ellipticals in $V-K$ and $U-V$
 (Bower, Lucey, \& Ellis 1992a,b) at the galaxy age $T_G=$12 Gyr.
 Changing $t_{gw}$ is equivalent to adjusting the 
 mean stellar metallicity of the
 galaxies, therefore we call this the 
 {\it metallicity sequence} of elliptical
 galaxy models.
 In this model, the mean stellar metallicity $\langle$[M/H]$\rangle$
 changes from 0.06 to $-$0.52
 over a six magnitude range from the
 brightest E model ($M_V=-23$ mag at $z=0$).
 The time until the onset of a galactic wind $t_{gw}$ 
 is always shorter than $\sim$0.5 Gyr, thus the star formation in
 elliptical galaxies is burst-like.
 The above model sequence is shown to reproduce the evolution of the
 {\it C-M} relation of elliptical galaxies in distant clusters
 in Kodama \& Arimoto (1997) and Kodama et al. (1998).

 To represent the photometric properties of disk galaxy bulges,
 we borrow the elliptical galaxy models.  Observational support
 for this includes the results of Mg$_2$ index
 analysis (Jablonka, Martin, \& Arimoto 1996).

\subsubsection{Disks}

\begin{table*}
\caption{Integrated colours of spiral galaxies.}
\label{tab:spiral}
\begin{center}
  \begin{tabular}{c|c|cccc|cccc|ccccc}
  \hline\hline
  & & \multicolumn{4}{c}{de Jong (1996)} & \multicolumn{4}{c}{RC3} &
  \multicolumn{5}{c}{Model}\\
  \hline
  Hubble type & B/T & B$-$V & V$-$R & V$-$I & V$-$K & U$-$V & B$-$V & V$-$R &
  V$-$I & U$-$V & B$-$V & V$-$R & V$-$I & V$-$K \\
  \hline
  Sa & 0.41 & 0.81 & 0.54 & 1.02 & 2.94 & 0.96 & 0.74 & 0.50 & 1.14 &
  0.85 & 0.76 & 0.53 & 1.17 & 3.02 \\
  Sb & 0.24 & 0.74 & 0.46 & 1.04 & 2.79 & 0.66 & 0.61 & 0.46 & 0.93 &
  0.63 & 0.66 & 0.49 & 1.07 & 2.84 \\
  Sc & 0.09 & 0.67 & 0.53 & 1.08 & 2.84 & 0.45 & 0.53 & 0.41 & 0.87 &
  0.45 & 0.57 & 0.43 & 0.95 & 2.61 \\
  Sd & 0.02 & 0.59 & 0.47 & 1.02 & 2.59 & 0.37 & 0.50 & 0.39 & 0.83 &
  0.36 & 0.52 & 0.41 & 0.89 & 2.47 \\
  Sm & 0.00 & 0.69 & 0.41 & 0.75 & 2.30 & 0.33 & 0.50 & 0.36 & 0.72 &
  0.33 & 0.50 & 0.40 & 0.87 & 2.42 \\
  \hline\hline
  \end{tabular}
\end{center}
\end{table*}

\begin{figure}
\begin{center}
  \leavevmode
  \epsfxsize 1.0\hsize
  \epsffile{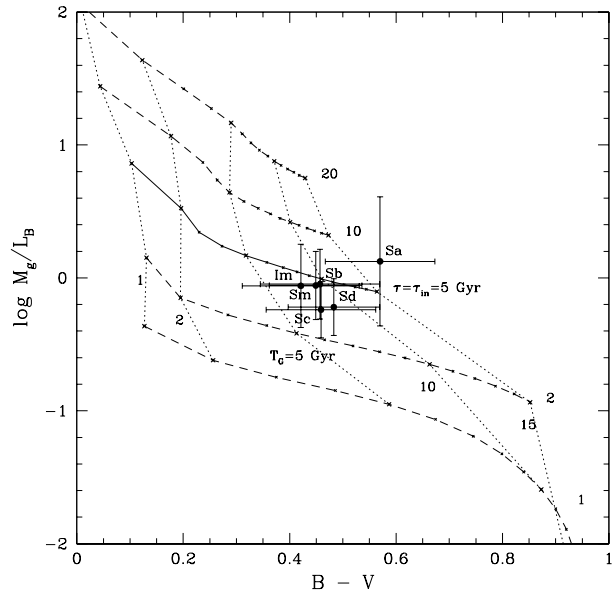}
\end{center}
\caption
{
 Galaxy disks. Gas mass per unit blue luminosity is plotted against $B-V$
 colours. The filled circles show the estimated disk colours
 of galaxies ranging between Sa and Im (see text).
 The solid line and the four
 dashed line represent the evolutionary models with different $\tau$ and
 $\tau_{in}$. The age $T_G$ is changed from 1 Gyr to 15 Gyr as indicated by
 crosses along the lines.
}
\label{fig:disk}
\end{figure}

 For the disk component, the IMF parameters are set to $x=1.35$,
 $m_{l}=0.1$, $m_{u}=60$ M$_{\odot}$, and longer time scales of star
 formation and gas infall: $\tau=\tau_{in}=5$ Gyr.
 The age of a galactic disk is fixed at 12 Gyr.
 The disk model time scales are chosen to reproduce
 the integrated $B-V$ colours and $M_g/L_B$ ratio of observed disks of
 various Hubble-types as shown in Fig.~\ref{fig:disk}
 (\cf Shimasaku \& Fukugita 1997).

 The $B-V$ colours of disks shown in Fig.~\ref{fig:disk} as a function
 of Hubble type are estimated from:
 \begin{itemize}
 \item the mean total $B-V$ colours as a function of Hubble type 
 (Buta \etal 1994), and
 \item subtraction of the bulge light by assuming a bulge
 colour $B-V=1.0$ and a bulge to total light ratio (B/T) in $B$-band
 (Simien \& de Vaucouleurs 1986).
 \end{itemize}
 The total gas masses normalised by $B$-band disk luminosity ($M_g/L_B$)
 as a function of Hubble type are estimated from:
 \begin{itemize}
 \item the mass of neutral atomic gas,
 calculated from the integrated hydrogen index \hi (Buta \etal 1994)
 and a conversion formula in {\it Third Reference Catalogue of Bright Galaxies}
 (RC3) given by de Vaucouleurs \etal (1991),
 \item the ratio of molecular to atomic gas H$_2/$\hi (Young \& Knezek 1989),
 \item a helium abundance correction of 25\%, and
 \item subtraction of the bulge light contribution to $L_B$.
 \end{itemize}

 Disk properties in Fig.~\ref{fig:disk}
 are well reproduced by $\tau=\tau_{in}=5$ Gyr model with an age $T_G=5-15$
 Gyr irrespective of the Hubble type.
 The model also reproduces the age-metallicity relation and the
 [O/Fe] \vs [Fe/H] diagram of the stars in our own galaxy (Kodama 1997).
 The constraint on the time scales $\tau$ and $\tau_{in}$ is weak
 because of the large observational errors and the permitted range
 could be from 2 to 8 Gyr (Fig.~\ref{fig:disk}).
 However, as will be shown in the next section (\S~2.2),
 this uncertainty will not cause problems for the purposes of 
 redshift determination
 because star forming timescale and B/T variations have degenerate effects.

 As an additional check of the validity of our models, the
 integrated colours of disk galaxies of different Hubble type 
 are compared in Table \ref{tab:spiral}.  
 The observational data are mean Hubble type colours 
 taken from de Jong (1996) and the RC3
 (Buta et al.\ 1994; Buta \& Williams 1995). Note that
 each galaxy type has large intrinsic colour dispersion, typically as much
 as $0.05-0.2$ mag in optical colours and $0.2-0.4$ mag in $V-K$.
 The data are compared to the model with appropriate $B$-band 
 B/T ratio (Simien \& de Vaucouleurs 1986).  It is clear that the 
 detailed trends of local galaxy colour with $B$-band B/T ratio are 
 well reproduced by our models. 

\subsection{Colour tracks \label{subsec:coltrack}}

\begin{figure*}
\begin{center}
  \leavevmode
  \epsfxsize 1.0\hsize
  \epsffile{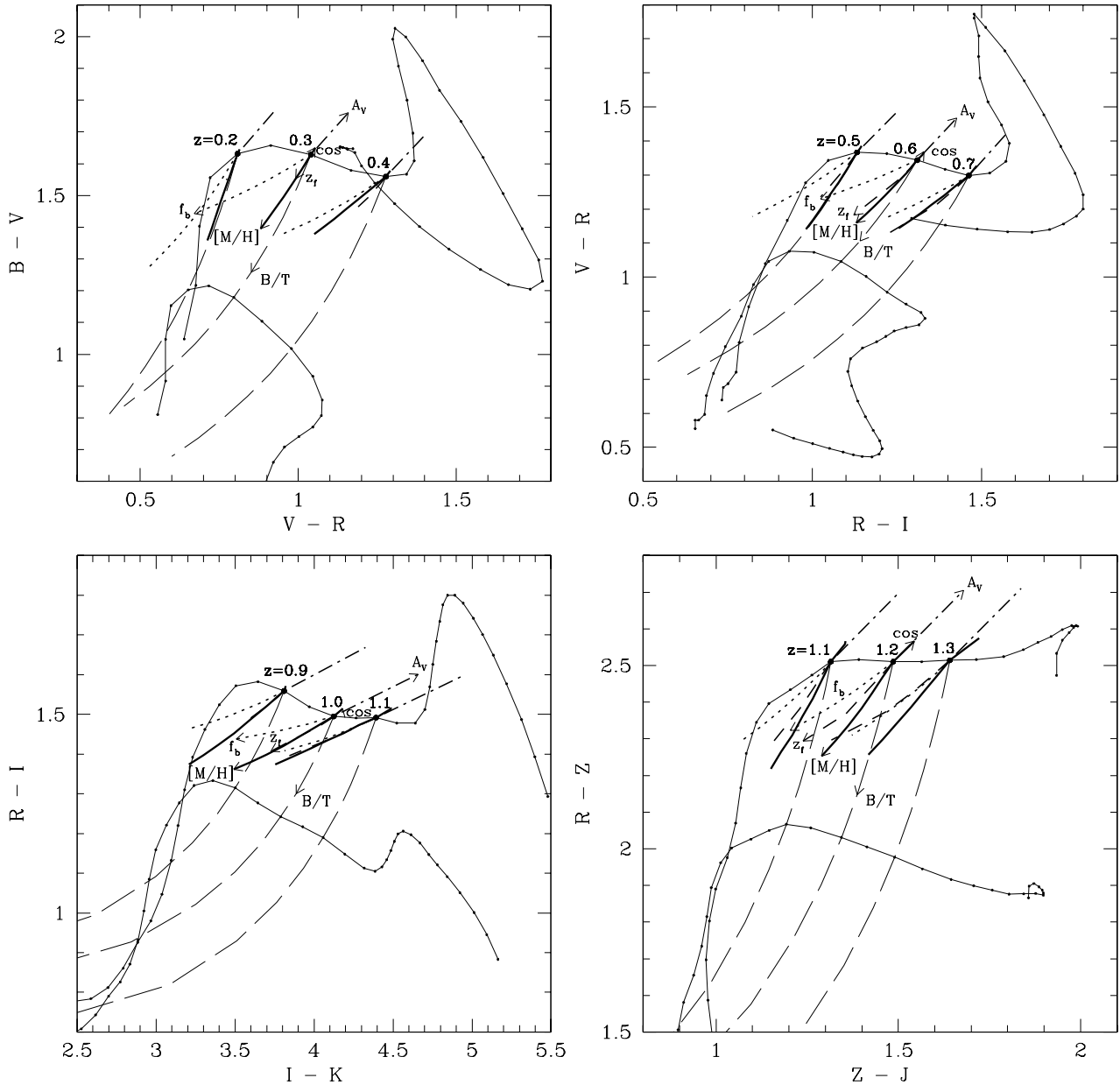}
\end{center}
\caption
{
 Colour-colour plots in the observer's frame.
 The two solid curves show the colour evolution of an old ($T_G=12$ Gyr,
 or $z_ f=4.5$), high metallicity ($\langle$[M/H]$\rangle$$=$0.06) E model
 (redder),
 and the B/T=0.5 model (bluer) which contains half of disk light in $B$-band
 at $z=0$.
 The redshift is changing from 0 to 2 in steps of 0.05, indicated by dots
 along the tracks.
 The six arrows shown from three redshift points along the track indicate the
 colour changes due to several different effects. See text for detail.}
\label{fig:twocolour}
\end{figure*}

 Following the models introduced above,
 we simulate the colour evolution of galaxies as a function of redshift
 for variety of star formation histories.

\begin{figure}
\begin{center}
  \leavevmode
  \epsfxsize 1.0\hsize
  \epsffile{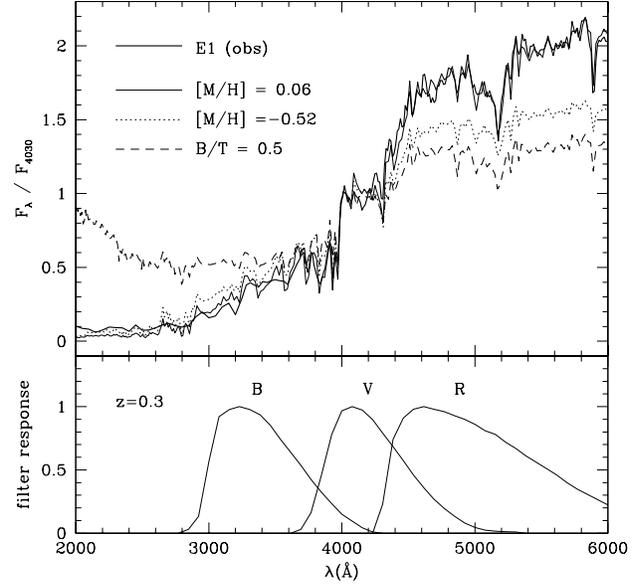}
\end{center}
\caption
{
The galaxy spectra at $z=0$ (upper panel).
The flux is normalised at 4030~{\AA}.
The thick solid line shows a giant elliptical E1 spectral template 
(Bica 1988) extended into the UV by the attachment of IUE spectra
(Arimoto 1996).
Three model spectra are superposed:  a high mean metallicity
model with $\langle$[M/H]$\rangle=0.06$ (thin solid line), a lower
metallicity model with $\langle$[M/H]$\rangle=-0.52$ (dotted line),
and a high metallicity bulge plus disk model with B/T=0.5 (dashed line).
See text for details of the models.
The lower panel shows the normalised response functions of standard
Johnson-Cousins' $B$,$V$ and $R$ filters, blueshifted
to correspond to those at $z=0.3$.
}
\label{fig:spectra}
\end{figure}

 The two solid curves in Fig.~\ref{fig:twocolour} show the colour
 evolution in the observer's frame for a E model with a high metallicity
 ($\langle$[M/H]$\rangle$=0.06),
 and a model which contains 50\% contribution of disk light in the $B$-band
 at $z=0$ (see below).
 The redshift is changed from 0 to 2 in steps of 0.05 as 
 indicated by the dots along the lines.
 Four different colour-colour plots 
 are shown, to cover a wide range in redshift, demonstrating that
 the most useful passbands for photometric redshift determination up to 
 redshifts of $\sim1.5$ typically bracket the 4000~{\AA} break:
 $0.25 \leq z \leq 0.4$ for $B-V$ \vs $V-R$ colours,
 $0.5 \leq z \leq 0.8$ for $V-R$ \vs $R-I$ colours,
 $0.9 \leq z \leq 1.15$ for $R-I$ \vs $I-K$ colours,
 and $1.0 \leq z \leq 1.5$ for $R-Z$ \vs $Z-J$ colours.
 In the above redshift ranges,
 the middle bands of each combination
 are passing through 4000~{\AA} break, the most prominent
 spectral feature in optical region, which plays an important role in
 redshift estimation.
 The horizontal colours redden rapidly with
 redshift while vertical colours stay nearly constant.
 At around $z=0.3$ for example, as shown in Fig.~\ref{fig:spectra},
 $V$ band is just on the 4000~{\AA} break and
 $B$ and $V$ bands are losing flux rapidly, 
 while the $R$ band flux is approximately constant as the redshift
 increases. 
 As a result $V-R$ gets redder while $B-V$ remains almost constant.
 On the other hand, $B-V$ is more sensitive to changes in stellar
 population than $V-R$.  Therefore, we can see that
 the effects caused by changes in redshift
 are almost perpendicular to those caused by changes in star formation
 history.

In Fig.~\ref{fig:twocolour}, six possible effects which change the 
colour evolution of a galaxy with redshift are considered.
The six arrows (indicated at three redshift points along the colour track)
show the change in colour of an old, high metallicity elliptical model 
($T_G=12$ Gyr and $\langle$[M/H]$\rangle$$=$0.06)
caused by the following effects.

\begin{enumerate}
\item {\it metallicity --- lower thick solid arrows:\ }
   The mean stellar metallicity of the E models is changed from
   $\langle$[M/H]$\rangle$$=$0.06 to $-$0.52.
   The galaxy age is fixed at 12 Gyr,
   corresponding to a formation redshift $z_f=4.5$.
\item {\it age --- dashed arrows:\ }
   The formation redshift of the E model is varied from $z_f=$ 4.5 to 1.0
   (4.5 to 2.0 for the $RIK$ and $RZJ$ diagrams),  corresponding
   to galaxy ages $T_G=$ 12.0 to 8.4 Gyr (12.0 to 10.5 Gyr).
   Metallicity is fixed at $\langle$[M/H]$\rangle$$=$0.06.
\item {\it disk component --- long dashed arrows:\ }
   As outlined earlier, the models deal with disk galaxies by 
   adding a star forming disk component onto an E model,
   representing the galaxy bulge.  The 
   $B$-band B/T ratio at $z=0$ is changed from 100\% to 0\%.
   Note that B/T=0.41, 0.24 and 0.09 corresponds to the Hubble
   types Sa, Sb and Sc, respectively
   (Simien \& de Vaucouleurs 1986). 
   Note also that B/T ratio is a rising function with redshift,
   since the bulge is getting brighter with redshift while the disk brightness
   remains roughly constant up to $z=1$.
   For example, the B/T ratio of 0.5 at $z=0$ actually increases to 0.8
   at $z=1$.
   We have also modelled the B/T sequence using
   [M/H]$_{\rm disk}=-1.3$ and [M/H]$_{\rm disk}=0$ by forcing all the disk
   stars to have
   the same metallicity instead of following the 
   chemical evolution of the disk.
   [M/H]$_{\rm disk}$ is found to have a negligible effect on the colour
   tracks of the B/T sequence, and is not considered hereafter.
   The time scales $\tau$ and $\tau_{in}$ can be also changed,
   but it is found to be similar in effect to changes in
   B/T ratio at fixed $\tau$ and $\tau_{in}$.
   This is because the colour of the B/T sequence at a given redshift
   is essentially determined by the ratio of the 
   current star formation rate and the total mass of the 
   galaxy.  This parameter can be adjusted
   either by changing the time scales of the disk or by changing the
   B/T ratio. 
   Thus the effects on the colour-colour diagrams of changing
   the timescales is only to shorten or extend the vectors of the
   B/T sequence for a given B/T ratio.
\item {\it recent star burst --- dotted arrows:\ }
   The possible effects of recent large-scale star formation are
   considered by the addition of a recent star burst to the E model.
   The burst population is assumed to be a simple stellar population (SSP)
   with solar metallicity.
   The arrow denotes the changes caused by a $T_b = 0.5$ Gyr old
   burst population corresponding to 10\% of the total stellar mass
   ($f_b=0.1$) at that redshift.
   The direction of the vector on
   the colour-colour diagrams depends on $T_b$, but unless $T_b$ is around 0.5
   ($\pm 0.3$) Gyr, the burst sequence follows either the B/T sequence or
   the age sequence closely. 
   Main-sequence turn-off stars in the burst population
   with age $\sim0.5$ Gyr have an effective temperature of about 10000 K and
   contribute significantly to the total 
   flux at rest wavelengths of 3000-4000~{\AA}, 
   creating aberrant colour changes in the colour-colour plots.
\item {\it reddening -- dash-dotted arrows:\ }
   The extinction effect due to internal dust is estimated
   by using the extinction curve given by Mathis (1990).
   The full arrows correspond to $A_V=0.5$ mag.
\item {\it cosmology --- upper thick solid arrows:\ }
   The colour tracks have a weak dependence on the adopted cosmology.
   Other sets of cosmological parameters are tested; \ie
   $H_0=65, \Omega_0=0.1, \Lambda_0=0.0$, 
   and $H_0=80, \Omega_0=0.2, \Lambda_0=0.8$. 
   The formation redshift $z_f$ is fixed to 4.5 in all cases.
   The full arrows show the colour change for the latter cosmology.
   The colour change from the former cosmology is smaller and along
   a similar vector.

\end{enumerate}
 
The age and metallicity sequences (i) and (ii) are almost indistinguishable
in all the colour-colour plots for ages $\gsim1$ Gyr.
This reflects the age-metallicity degeneracy inherent in old stellar
populations (Worthey 1994).
However, this degeneracy actually improves the 
prospects for the determination of photometric redshifts, as the
effects of age and metallicity are quite distinct, given the right 
choice of passbands, to the effects of changing galaxy redshift.
In addition, it is clear from Fig. \ref{fig:twocolour} that 
changes in assumed cosmology and interstellar reddening also 
have colour effects similar to age and metallicity, with an opposite
sense.  
As a result, E-type galaxies at a given redshift should populate in a 
restricted area on the colour-colour diagram (almost a single line)
characteristic of that redshift, irrespective of its stellar population,
regardless of interstellar reddening, and whichever cosmology is assumed.
This means that it is possible to assign redshifts to old stellar 
populations without prior knowledge of galaxy properties.

However, the colour changes caused by the B/T sequence (iii) 
are not entirely degenerate with those due to age and metallicity
on the $RIK$ and $RZJ$ diagrams (Fig.~\ref{fig:twocolour}), 
due to the presence of on-going star formation.  This on-going star 
formation causes a bluer \eg $R-I$ colour for a given $I-K$ colour
than either the effects of age and metallicity for $z \sim 1$ galaxies.
Recent large bursts of star formation of age $\sim0.5$ Gyr (iv) also lead
to effects distinct from those of age and metallicity, and those of 
changing the B/T ratio.  

This can lead to considerable uncertainty in the estimation of galaxy
redshift, as a given set of colours, on the basis of the colour-colour
plots presented in Fig.~\ref{fig:twocolour}, will be consistent with 
a wide range of redshifts, depending on how the colours are explained
by our model; \eg by changing the B/T ratio, or the metallicity of the
galaxy template.  However, if a passband with a short rest-frame 
wavelength is used, it is possible to discriminate the presence of 
young stellar populations photometrically, leading to a less ambiguous 
determination of redshift, and some information on the star formation
history of the galaxy.  This is illustrated in the upper
half of Table~\ref{tab:degeneracy}
(see also the lower left panel of Fig.~\ref{fig:twocolour}), where
we compare 3 galaxy templates with very similar red colours 
($R-I \simeq 1.35$, $I-K \simeq 3.48$) which present
very distinct colours in bluer passbands, allowing
relatively easy discrimination between these possibilities.
Another degeneracy apparent in Fig.~\ref{fig:twocolour} in redder passbands 
is that between high redshift, low B/T ratio galaxies, and lower redshift
early-type galaxies.  This, again, is illustrated in the lower
half of Table~\ref{tab:degeneracy}
where we again see that bluer passbands allow easy splitting of
this degeneracy.

\begin{table*}
  \caption{Colour degeneracies.}
  \begin{tabular}{cccccccc}
  \hline
  $z$ & B/T & $\langle$[M/H]$\rangle$ & $f_b$(\%) &
  $B-R$ & $V-R$ & $R-I$ & $I-K$ \\
  \hline
  0.9 & 0.5 & 0.06    & 0  & 1.32 & 0.86 & 1.32 & 3.50 \\
  1.0 & 1.0 & $-0.52$ & 0  & 2.57 & 1.32 & 1.36 & 3.49 \\
  1.1 & 1.0 & 0.06    & 20 & 1.64 & 1.09 & 1.38 & 3.46 \\
  \hline
  0.5 & 1.0 & 0.06    & 0  & 2.98 & 1.37 & 1.13 & 3.10 \\
  0.7 & 0.4 & 0.06    & 0  & 1.48 & 0.89 & 1.15 & 2.98 \\
  0.9 & 0.2 & 0.06    & 0  & 0.80 & 0.57 & 1.09 & 3.00 \\
  \hline
  \label{tab:degeneracy}
  \end{tabular}
\end{table*}

Another point to note from Fig.~\ref{fig:twocolour} is that when a
particular colour pair is selected to allow the accurate estimation of
redshifts within a certain redshift range, this colour pair also
provides a means of rejecting galaxies (particularly higher B/T objects)
that lie outside this optimal redshift range (although the estimated
redshifts will obviously be much less accurate for these objects).
Problems will occur for much higher redshift objects, and objects with
a small B/T ratio, as discussed above.

Despite the demonstrated utility of the bluer passbands
in `breaking' degeneracies between galaxies which look identical
in red passbands,
we aim to use little of the colour information shortwards of 2500~{\AA}.
Primarily, this is because the model spectra are ill-constrained for short UV
wavelengths in both elliptical and star forming galaxies because
of the effects of the UV-upturn (an anomalous rise in flux towards
short UV wavelengths, observed in nearby giant ellipticals;
eg., Burstein et al.\ 1988) and the uncertain effects of dust
extinction (White, Keel, \& Conselice 1996).
The source of the UV-upturn is still poorly understood, and the model
predictions for its source and effects are still uncertain.
If the UV-upturn comes from hot young stars, this population is
actually considered in this model by superposing only a small fraction of
on-going star formation onto the passively evolving ellipticals.
If, however, the source of the UV-upturn is hot horizontal branch stars,
then it is necessary to fine tune the mass loss parameter along the red giant
branch to reproduce the UV-upturn (\cf Yi, Demarque, \& Oemler, 1997).
Even if this were the case, such hot horizontal branch stars
would disappear at high redshift ($z \gsim 1$), which is our main
region of interest, since the envelope mass of a horizontal
branch star gets larger as the
mass of a main sequence turn off star gets larger with look back time.

An additional source of uncertainty in our models, especially in the 
UV, is the neglect of the effects of dust extinction on the colours
of the stellar populations incorporating on-going star formation.
This would at first appear to be a serious handicap, as disk-dominated 
galaxies clearly contain significant amounts of dust, especially in
the spiral arms, where $B$ band extinction $A_B \sim 1$ mag 
(White, Keel, \& Conselice 1996, Berlind et al.\ 1997).  However,
by inspection of Fig.~\ref{fig:twocolour}, it is clear that the colour
changes at rest-frame optical wavelengths are equivalent to 
increasing B/T ratio, age {\it or} metallicity, meaning that relatively
large uncertainties in dust reddening can be accommodated by changes
in other galaxy parameters to compensate for these errors.
This is also demonstrated in Table \ref{tab:spiral}, where it is apparent
that our models can accurately reproduce the colours of galaxies 
with ongoing star formation.  This situation is unlikely to hold in 
the UV, however, as prescriptions for the dust extinction law 
start to diverge at these short wavelengths (Calzetti, Kinney,
\& Storchi-Bergmann 1994).

Both the UV-upturn and the uncertain effects of dust reddening in the far-UV
lead us to place little confidence in our model UV colours.  We should
therefore avoid this spectral region if possible.

In addition, it should be noted that we neglect emission from star
forming galaxies, such as the commonly observed [\oii], [\oiii] and
Balmer features at locally (Kennicutt 1992) and at high redshifts
(Hammer et al.\ 1997).  This should not present a major
problem, as the effects of line emission on broad band photometry is
not large: A line width with an equivalent width of 20 {\AA} in
emission would cause only $\sim 0.02$ mag of brightening in the broad
band magnitude.

Finally, we note that although most of redshift range below 1.5 can be
covered by the standard Johnson-Cousins system including $Z$ band,
there are some particular redshift range where we have larger errors
in the estimated redshifts; ie. $z<0.25$, $0.4<z<0.5$, and $0.8<z<0.9$.
At these redshift ranges, the effect of changing redshift on colours is
hard to be distinguished from that of changing stellar population
(Fig.~\ref{fig:twocolour}).
If we want to handle clusters in these redshift ranges with better precision,
we need to use passbands in other photometric systems which properly bracket
the 4000 {\AA} break at the cluster redshifts.

\section{Bayesian classification}

\subsection{Basic scheme}

A Bayesian approach allows us to incorporate our existing knowledge of
galaxy populations, and thus to proportionally weight the areas of
parameter space that we search. The Bayesian probability of a particular
galaxy having a redshift $z$ and bulge to total luminosity ratio
B/T is given by the equation:
\begin{equation}
P_{\rm{Gal}}(z,{\rm B/T}) = P_1(z,{\rm B/T}{\mid}m_B) P_2(z,{\rm B/T}),
\label{eqn:probability}
\end{equation}
where $P_1(z,{\rm B/T}{\mid}m_B)$ is the probability of a given galaxy of
aparent magnitude $m_B$ having a redshift $z$ and bulge to total luminosity
ratio B/T, and $P_2(z,{\rm B/T})$ is the probability of a given galaxy
reproducing the observed galaxy colours.  We first deal with the
evaluation of $P_2(z,{\rm B/T})$, \ie the probability of a given model galaxy
reproducing the observed galaxy colours.

The basic philosophical approach used for 
this redshift estimator is the comparison of a galaxy's
location on a colour-colour plot and a finely-spaced 
grid of models superimposed on
that plot to estimate the properties of that galaxy.  
The magnitudes of the observed galaxy are made into colours, and the
errors in the colours used to make up a covariance matrix, describing
the sizes of the colour errors, and their relationships.
Then, for all of the model galaxy colours, the difference between them
and the observed colours are calculated.  Under the assumption that the
photometric errors follow a Gaussian distribution, the probability that the 
model describes the galaxy colours adequately ($P_2(z,{\rm B/T})$) is given by:
\begin{equation}
P_2(z,{\rm B/T}) = 1/{[(2\pi)^n {\mid}\mbox{\boldmath $C$}{\mid}]^{1/2}} \,
\exp \{-1/2 (\mbox{\boldmath $u^T C^{-1} u$})\},
\end{equation}
where $n$ is the number of colours used, {\boldmath $u$} is the vector of
differences between the model galaxy colours and the observed colours,
{\boldmath $C$} is the covariance matrix of those colours, and 
${\mid}$\mbox{\boldmath $C$}${\mid}$ is the
determinant of the covariance matrix.
The diagonal elements of {\boldmath $C$} correspond to the variance in
the individual colours.  The off-diagonal elements correspond to the
variance of any passbands in common between two colours, with the
appropriate sign (which indicates whether the errors in a given
passband affect the colours in the same or an opposite sense).
Since we require galaxies to have small ($\lsim$ 0.1 mag) 
photometric errors in order to
reliably determine their redshift, using a Gaussian rather than a
log-normal is an adequate representation of the error distribution in
each band.  This assumption considerably shortens the probability
computation time.
If it is impossible to find a satisfactory match to the galaxy colours,
the galaxy is omitted from further consideration.

This procedure gives the probability that an observed galaxy is
adequately described by a given model galaxy.
This product is evaluated for a large number of plausible model galaxies
with a range of B/T ratio and redshift. The other effects, such as age and
metallicity of model galaxies are considered later in \S~3.3.
These evaluations makes up a `probability map' on the plane of B/T ratio
and redshift.
In order for us to obtain the final probability,
$P_{\rm{Gal}}(z,{\rm B/T})$ of the galaxy having a particular $z$
and B/T, it is necessary to make up a prior distribution, given what
is already known about galaxy populations.

\subsection{Prior distribution}

The quantity $P_1(z,{\rm B/T}{\mid}m_B)$ is our prior distribution. 
The effect of the prior is to modulate the redshift estimates
provided by the colour analysis by using magnitude information.
Note that it makes little difference to the redshift
estimate (well within the error bars of the redshift determination)
{\it unless} there is a degeneracy, and the colour of two or more
models satisfy the observational constraints equally well at different
redshifts.
In that case, it is designed to discover which one of these options is more
likely to be observed, and weights the `probability map' accordingly.

In forming the prior distribution,
we need to know the type dependent luminosity function (LF)
$\Phi(m_B,{\rm B/T})$ and the volume element $dV/dzd\Omega$. 
Using these two elements, the prior is given as follows:
\begin{equation}
P_1(z,{\rm B/T}{\mid}m_B) = dV/dzd\Omega \, \Phi(m_B,z,{\rm B/T}).
\end{equation} 
These two parts are treated separately below.

\subsubsection{The local luminosity function}

\begin{table}
  \caption{Adopted LF parameters.}
  \begin{tabular}{llccc}
    \hline
    Type & B/T & $M_B^*$ & $\alpha$ & $\Phi^*$ ($\times 10^{-3}$ Mpc$^{-3}$) \\
    \hline
    E                      & 1.0  & $-$20.74 & $-$0.85 & 0.188 \\
    S0                     & 0.6  & $-$20.25 & $-$0.94 & 0.950 \\
    Sa--Sb & 0.33 & $-$20.23 & $-$0.58 & 1.088 \\
    Sc--Im & 0.0  & $-$20.32 & $-$1.07 & 0.625 \\
    \hline
  \label{tab:LF}
  \end{tabular}
  \medskip
\end{table}

In order to get $\Phi(m_B,{\rm B/T})$ we use the local, type-dependent
luminosity function (LF). 
However, there remains considerable uncertainty in the type-dependent
LF, as the splitting into morphological types is carried out in a
number of different ways, and the faint end slopes differ considerably
between different studies (Marzke et al.\ 1998, Bromley et al.\ 1997,
Marzke et al.\ 1994, Bingelli, Sandage, \& Tammann 1988).
We chose to adopt a variant of 
Marzke et al.'s (1994) determination in the Schechter (1976)
form, which is parameterised by
$\Phi^*$, $\alpha$ and $M_B^*$.  The parameters for the
observed local luminosity function are summarised
in Table~\ref{tab:LF} as a function of the Hubble type.
To connect between the Hubble types and B/T ratio,
we use Simien \& de Vaucouleurs (1986).
The characteristic magnitude of the LF, $M_B^*({\rm B/T})$, corresponds to
the apparent magnitude $m_B^*(z,{\rm B/T})$ at a redshift $z$ as:
\begin{equation}
m_B^*(z,{\rm B/T}) = M_B^*({\rm B/T}) + DM(z) + \Delta M_B(z,{\rm B/T}),
\label{eqn:deltab}
\end{equation}
where $DM$ is the distance modulus at redshift $z$ in the adopted cosmology.
$\Delta M_B$ is the absolute magnitude change in $B$-band in the
observer's frame due to the luminosity evolution and the shift of the
wavelength shortwards with redshift, and is taken from the model.
In this way, we finally obtain the LF in apparent magnitude $m_B$ as a function
of redshift and B/T ratio:
\begin{eqnarray}
\lefteqn{\Phi(m_B,z,{\rm B/T}) = } \nonumber \\
& & 0.92 \Phi^* e^{\{-0.92(\alpha+1)
(m_B-m_B^*)-\exp(-0.92(m_B-m_B^*))\}}.
\label{eqn:LF}
\end{eqnarray} 

If the observed galaxy lacks $B$-band data, we use a prior in the band nearlest
to $B$. In this case, we make up the local LFs in the alternative band
by shifting the $B$-band LFs using model colours of each type.

\subsubsection{Volume element}

The other essential ingredient of the prior is the volume element
$dV/dzd\Omega$.  The formula for the volume element as a function of
redshift was taken from Carroll, Press, \& Turner (1992), with the
addition of some factors of $c$ to satisfy dimensionality
considerations, and allows
variations in $\Omega_0$, $H_0$ and the inclusion of the
cosmological constant via the term $\Lambda_0$:  
\begin{equation}
dV/dzd\Omega = \frac{d_M^2}{\{1+\Omega_k(H_0 d_M / c)^2\}^{1/2}} \, 
  \frac{d(d_M)}{dz}, 
\end{equation}
where, $\Omega_k$ is given by $\Omega_k=-\frac{kc^2}{R_0^2 H_0^2}$, and
$R_0$ is the scale factor of the universe and $k$ is the curvature of
the universe.  The quantity $d_M$ is the proper motion distance, and in
this case is given by:
\begin{equation}
d_M(z) = \frac{c}{H_0{\mid}\Omega_k{\mid}^{1/2}}\, \rm{sinn}
  \,({\mid}\Omega_k{\mid}^{1/2} {\cal F}),
\end{equation}
where `sinn' is a function that equals sinh in an open universe, sin in
a closed universe, and disappears in a critical universe, and
${\cal F}$ is given by:
\begin{equation}
  {\cal F} =  \int_0^z 
    [(1+z')^2(1+\Omega_0 z') -
    z'(2+z')\Lambda_0]^{-1/2}dz',
\end{equation}
which must be integrated numerically for most non-trivial cosmologies. 

\subsubsection{Comparison with observation}

\begin{figure}
\begin{center}
  \leavevmode
  \epsfxsize 1.0\hsize
  \epsffile{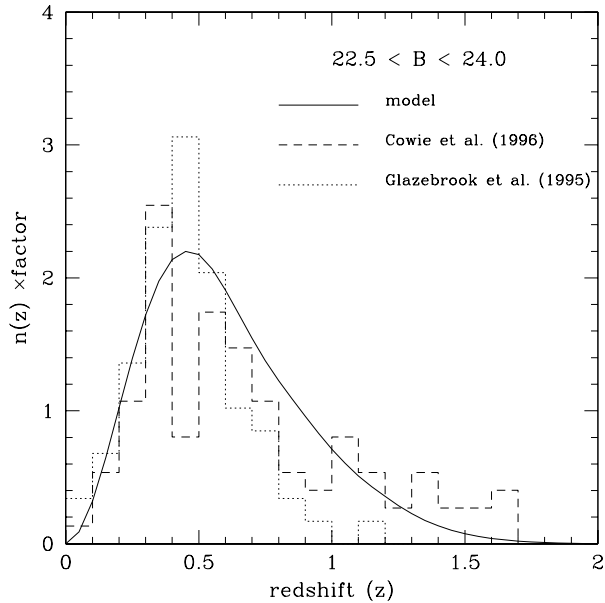}
\end{center}
\caption
{
 The observed $n(z)$ distributions for a sample of galaxies in the
 magnitude range $22.5<m_B<24.0$ from Glazebrook et al. (1995) and Cowie et al.
 (1996) are plotted against the expectation for the $n(z)$ distribution
 calculated from our prior distribution.
 The overall shape is in adequate agreement with the observational data.}
\label{fig:prior}
\end{figure}

The prior was used to calculate the $n(z)$ distribution within a magnitude
range $22.5<m_B<24.0$.  This calculated distribution was then compared to
the observed redshift distribution of galaxies within the same magnitude
range given by Glazebrook et al. (1995) and Cowie et al. (1996).
The comparison is shown in Fig. \ref{fig:prior}.
It is clear that the prior reproduces the overall form of the observed $n(z)$
diagram.

\subsubsection{The effect of the prior}

\begin{figure}
\begin{center}
  \leavevmode
  \epsfxsize 1.0\hsize
  \epsffile{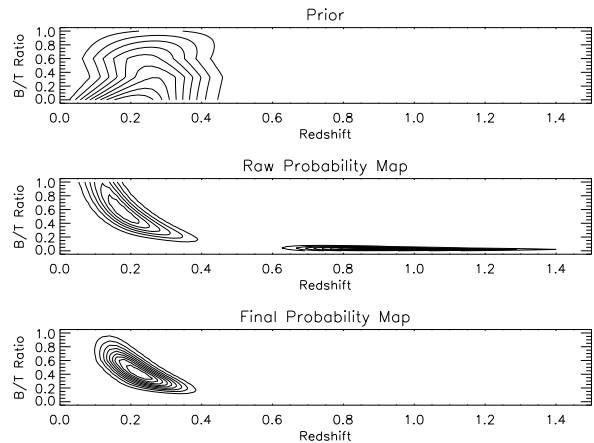}
\end{center}
\caption
{
Probability maps for a given galaxy with $m_V=20.3$.
The contours indicate linear probabilities in steps of 0.1 from the maximum.
Top panel shows the prior distribution ($P_1$), while the middle panel
indicates the probability map from the colour information only ($P_2$).
The bottom one is the final combined probability map.
The prior rejects the high redshift solution (which is permitted, when
considering only the galaxy's colours) based on the brighteness of the galaxy.
}
\label{fig:map}
\end{figure}

It should be noted that the prior distribution is quite model-dependent,
because the type-dependent local LF is ill-constrained, and because the
detailed type-dependent spectral evolution is poorly understood.
Also, here we make two assumptions: ie., that there is no number
evolution of galaxies, and that there is no size dependent luminosity
evolution, that is that galaxies with similar B/T ratios have the same colours
at all redshifts, regardless of their total luminosity.
These assumptions and ingredients may be inadequate to describe the
real universe, especially in the context of a hierarchical clustering
universe.
However, these uncertainties are not so important, because the estimated
redshift is essentially determined by the colour term
($P_2$ in Eq.~\ref{eqn:probability}), and the prior ($P_1$) is used
supplememtarily.
The prior becomes important when the solution from the colour term splits into
multiple redshift ranges. In such a case, the prior works to avoid
unreasonable solutions of redshift for a given apparent magnitude.
This situation is illustrated in Fig.~\ref{fig:map}.
The figure shows an example of the probability maps of a given galaxy.
The colour term gives two solutions, one at low redshift ($z\sim0.2$) and the
other at high redshift ($z\sim1.0$), but the prior rejects the solution with
higher redshift based on the brightness of the galaxy.

We also tested the effect of local LF on the final estimated redshift
through the prior by changing the LF paramters listed in Table~\ref{tab:LF}.
We shifted $M_B^{*}$ by $\pm0.5$ magnitude for all types, resulting in
a shift of the redshift peak of the n(z) distribution in
Fig.\ref{fig:prior} by $\mp0.1$, and we tried fixing
the faint end slope $\alpha$
at $-1$ for all types.
In all cases, however, the change in the final estimated redshifts
was well within the estimated redshift errors.
This experiment shows our method is robust to uncertainties
in the prior estimation.

\subsection{The models included in the classifier}

The estimates of redshift and galaxy type will depend quite
sensitively on the detailed choice of model galaxy template.
In \S~2.2, we investigated the various effects on the galaxy colours,
namely the effects of age, metallicity, disk light addition, recent star
bursts, reddening, and cosmology.
However, many of these effects were found to be degenerate with one another.
In these models, effects due to age, metallicity, reddening and changes 
in cosmology are particularly degenerate. 
Therefore, by including only the metallicity effect explicitly in the
classifier, we also cover the other three effects at the same time
as they behave just like metallicity variations.
To this aim, we considered four B/T sequences with different bulge
metallicities.
Each sequence gives different probability map on the redshift and B/T
plane, and then they are combined into a single `probability map'
by performing a mean of the separate maps.  This mean is in essence a
weighted mean, as the most plausible metallicity for the model galaxies
will have the best match to the colours.
In this way, we take into account the metallicity effects explicitly, and
hence the other degenerate effects.

The remaining significant 
effect which is not covered is that of a recent burst of star formation.
As seen in Fig.~\ref{fig:twocolour} large amounts of relatively recent star
formation can make a galaxy look as if it has a lower redshift than it
actually does, if the colours are interpreted as being entirely due to
redshift, B/T and metallicity effects.
They might be simply assigned significantly underestimated redshifts.
However, as already shown, unless burst strength $f_b$ is as high as
$>15\%$ and the burst age $T_b \sim 0.5$~Gyr, the burst population
colour change is similar to those resulting from other effects.
These populations should be short lived and are expected to be rare.
Therefore omitting the recent star burst model will not affect our
global redshift estimation.
We should note, however, this could be a problem if significant fraction of
cluster members would be strongly affected by a recent burst, due to
cluster-cluster merging for example.
In such a case, we would need to include the extra set of models of recent
star burst to correctly estimate redshifts, although it would lead to greater
estimation errors.

\subsection{Error estimates}
  \label{sec:error}

At this stage, the redshift and galaxy type estimates are in the form
of the `probability map' $P_{\rm{Gal}}(z,{\rm B/T})$.  However, an
estimate of a given galaxy's redshift and type is often required for
\eg comparison with real redshifts, or cluster member identification.
Best estimates for redshift ($z_{\rm estimated}^{\rm best}$)
and effective 1$\sigma$ confidence intervals ($z_{\rm estimated}^{\rm min}$,
$z_{\rm estimated}^{\rm max}$) are
obtained by taking the $P_{\rm{Cum}}(z) = 0.5$ and the $P_{\rm{Cum}}(z)
= [0.16,0.84]$ intervals, respectively, of the cumulative distribution:
\begin{equation}
  P_{\rm Cum}(z) = \int_{0}^{z}dz' \int_{{\rm B/T}=0}^{{\rm B/T}=1} 
                     d({\rm B/T}) \, P_{\rm Gal}(z',{\rm B/T}).
\end{equation} 

These error estimates depend on the estimated photometric errors:
through the use of the covariance matrix {\boldmath $C$}, 
the error estimates in
the photometry are explicitly included in the determination of
$P_2(z,{\rm B/T})$.  Large uncertainty in the colours 
propagates through into larger uncertainties in the $(z,{\rm B/T})$
combinations capable of adequately reproducing the observed colours.
It is important to know how much photometric accuracy we need to achieve
the error in redshift within the expected bound, as it will certainly constrain
the accuracy of any future applications of this method.

\begin{figure*}
\begin{center}
  \leavevmode
  \epsfxsize 1.0\hsize
  \epsffile{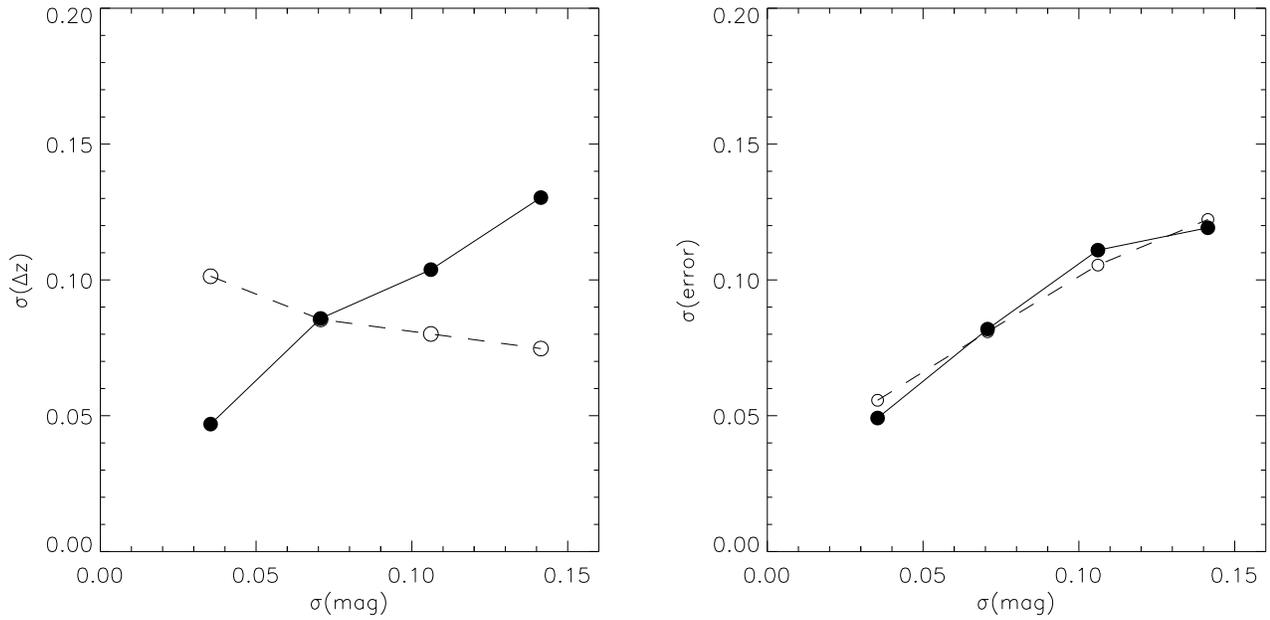}
\end{center}
\caption
{
The effects of photometric error on the quality of redshift estimation.
Simulations with 100 galaxies using the $B$, $V$, $R$, $I$ and $K$
passbands were used to assess the effects of
changing photometric quality in magnitude in all bands (solid lines).
The RMS difference between the best estimate and the real redshift
($\sigma(\Delta z)$) is plotted against the Gaussian photometric error
($\sigma(mag)$) in the left panel, while the RMS
redshift error ($\sigma(error)$) is plotted in the right panel.
The effects of under- or over-estimation of the photometric error were
also assessed by keeping the real photometric error fixed at 0.071 mag,
and varying the estimated photometric errors going into the covariance
matrix {\boldmath $C$} (dashed lines).
}
\label{fig:error}
\end{figure*}

To produce redshift error estimates as a function of photometric accuracy,
simulations using 100 galaxies with $K < 20$, chosen at random in the range of
$0<$B/T$<1$ and $0.2 \leq z \leq 1.8$ were undertaken.
The bulge metallicity $\langle$[M/H]$\rangle_{\rm bulge}$ was chosen randomly
between 0.061 and $-$0.523.
After allocating the model magnitude in each passband for each galaxy,
Gaussian photometric errors were then applied.
We used these magnitudes and photometric errors of the simulated galaxies
in the course of the redshift estimation.
Here $BVRIK$ passbands were used.
The quantities
$\sigma(\Delta z)$ and $\sigma(error)$, corresponding to the
root mean square (RMS) of real and estimated redshift error respectively,
were then plotted in Fig.~\ref{fig:error} where:
\begin{equation}
\Delta z = z_{\rm estimated}^{\rm best}-z_{\rm real},
\end{equation}
\begin{equation}
error = (z_{\rm estimated}^{\rm max}-z_{\rm estimated}^{\rm min})/2,
\end{equation}
\begin{equation}
\sigma(\Delta z) = \sqrt{\overline{(\Delta z)^2}},
\end{equation}
\begin{equation}
\sigma(error) = \sqrt{\overline{error^2}}.
\end{equation}
As seen from the solid lines, both $\sigma(\Delta z)$ and
$\sigma(error)$ increase with photometric error.
Importantly, even the photometric error is as bad as 0.15 magnitudes
in all bands, the average redshift error $\sigma(\Delta z)$ is still
kept smaller than 0.1. As for the estimated redshift error
$\sigma(error)$, it is roughly comparable to the photometric error.

Next we consider the effect of mis-estimation of the photometric error
on the redshift estimation error.
It is possible that if the errors are under- or
over-estimated, the redshift estimator will 
make the distribution of likely $(z,{\rm B/T})$ too broad or
multiply-peaked, reducing the accuracy of the redshift estimate.
Therefore, it is important to test the effects
that uncertainty in the determination of the errors can have on 
the redshift estimate accuracy. 
We realise this situation by fixing the real photometric error at 0.071
mag and varying the estimated photometric error 
going into the covariance matrix {\boldmath $C$}.
The result is shown by the dashed lines in Fig.~\ref{fig:error}.
It is clear that the under- or over-estimation of the photometric
errors has little, if any, effect on the quality of redshift estimation
$\sigma(\Delta z)$.
Errors in the determination of the photometric quality,
do however have a marked effect on the \emph{estimated} quality of the
redshift determination, given by $\sigma(error)$.  It is clear,
therefore that it is important to be careful in the estimation of the
photometric errors in order to estimate the quality of the redshift
determination effectively.

\begin{table}
  \caption{Passband choice for random galaxy simulations. \label{tab:band1}}
  \begin{tabular}{ccccc}
  \hline
  & \multicolumn{2}{c}{all} & \multicolumn{2}{c}{E/S0}\\
  passbands & $\sigma(\Delta z)$ & $\sigma(error)$ & $\sigma(\Delta z)$ & $\sigma(error)$\\
  \hline
   $BVRIK$ & 0.065 & 0.076 & 0.060 & 0.074 \\
   $RIK$   & 0.127 & 0.135 & 0.121 & 0.129 \\
   $BVRI$  & 0.178 & 0.199 & 0.176 & 0.210 \\
    \hline
  \end{tabular}
  \medskip
\end{table}

The quality of the redshift estimation also sensitively 
depends on the passband choice.
We investigated three sets of passbands for the randomly generated galaxies
with 0.071 mag photometric errors, and the results are summarised in
Table~\ref{tab:band1}.
With $RIK$ passbands, the result is about factor of two worse than the $BVRIK$
case. This is because it gets harder to disentangle the colour degeneracies
between lower redshift early-types and higher redshift late-types without
using bluer colours $B$ and $V$.
If, instead, we do not use $K$-band colours, the quality of the redshift
estimation worsens, as high redshift galaxies with $z>0.8-1.0$ no 
longer have a passband longwards of the 4000~{\AA} break.  It is therefore
important to choose the passbands carefully for photometric redshift
estimation according to redshift ranges under consideration and the depth
of the photometric sample.

\section{Testing}

In this section, we focus on testing our method using photometry for
galaxies with known spectroscopic redshifts.  Because we wish to focus
on the recovery of high redshift clusters at $z\gsim1.0$, it would be
best to test with an extensive dataset for a real high redshift
cluster.  
 However, such data is not available at the moment,
 since we need both multi colour photometry covering the 4000~{\AA} break
 (at least 3$-$4 bands)
 and spectroscopically determined redshifts for individual galaxies.
Therefore, we have decided to test our method with 
two independent sets of data:
a well-studied cluster Abell~370 at
$z=0.374$, and the Hubble Deep Field.

\subsection{Abell 370}

\begin{figure*}
\begin{center}
  \leavevmode
  \epsfxsize 1.0\hsize
  \epsffile{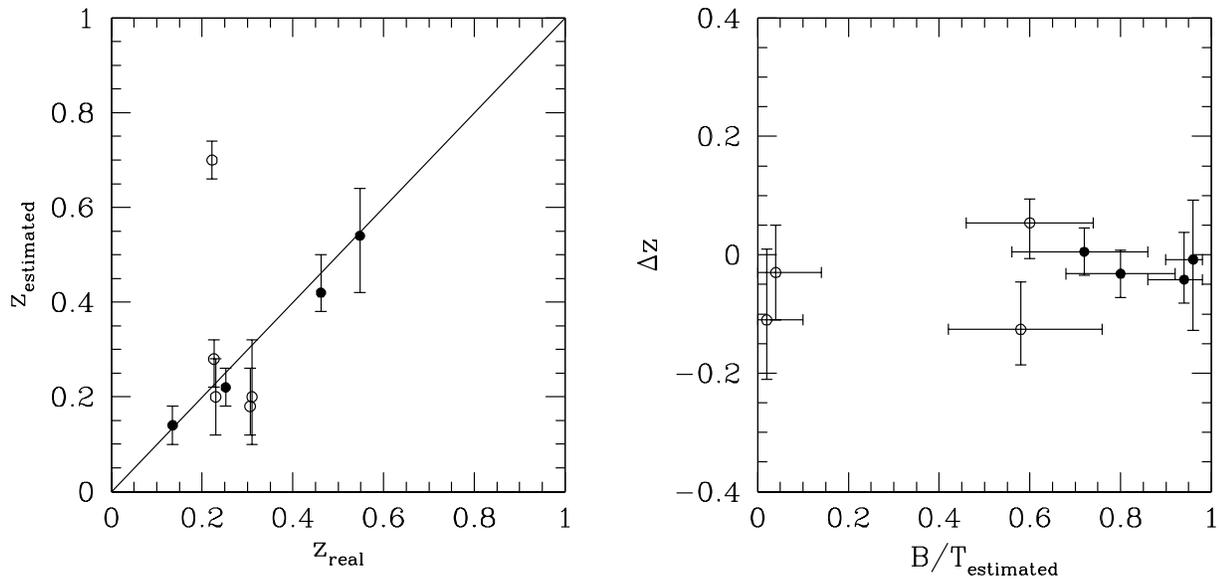}
\end{center}
\vspace{-9cm}
\caption
{
Field galaxies in the Abell 370 cluster field.
Estimated redshift vs. spectroscopically determined
redshift (left), and redshift error vs. estimated B/T ratio (right).
Filled circles indicate E/S0 galaxies,
while open circles indicate disk galaxies and those without morphological
information.
}
\label{fig:a370_field}
\end{figure*}

\begin{figure*}
\begin{center}
  \leavevmode
  \epsfxsize 1.0\hsize
  \epsffile{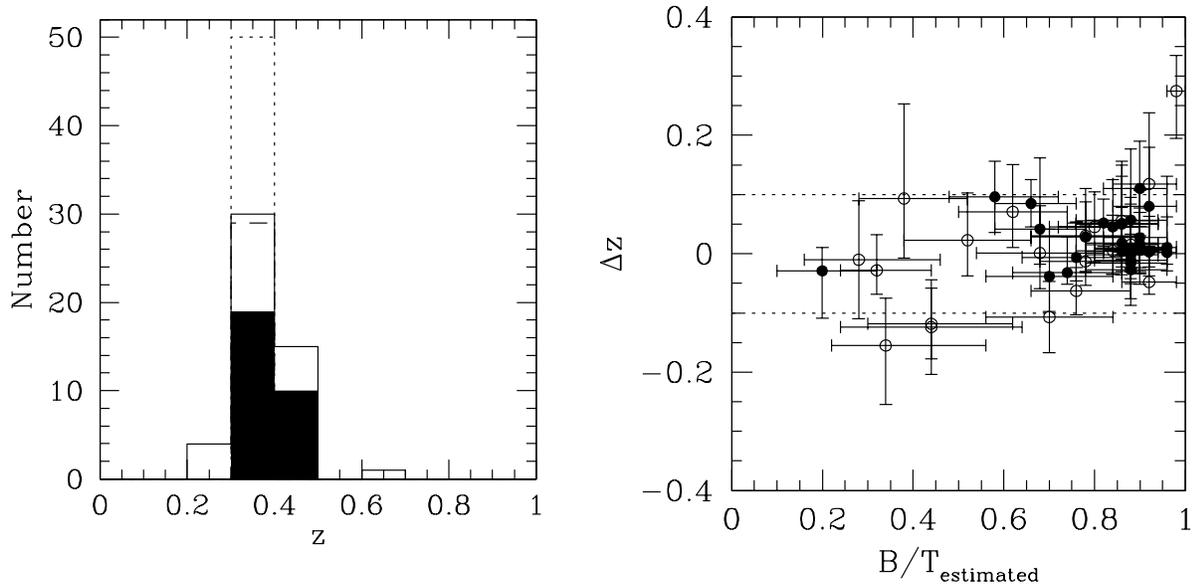}
\end{center}
\vspace{-9cm}
\caption
{
Cluster members of Abell 370.
Distribution of the estimated redshifts (left) and redshift error vs.
estimated B/T ratio (right).
The filled histogram and filled circles indicate E/S0 galaxies,
while the open histogram and open circles indicate 
disk galaxies and those without
morphological information.
The dotted and dashed histograms show the real redshifts
of all types and E/S0's, respectively.
The two dotted lines (right) show the region of $|\Delta z|<0.1$ where
galaxies are taken to be cluster members.
}
\label{fig:a370_cluster}
\end{figure*}

 The photometric data are taken from Pickles \& van der Kruit (1991).
 We use their $BVRI$ photometry, using a 7 arcsec aperture.
 For our use, we selected 59 galaxies which have spectroscopically
 determined redshifts. The redshifts are mainly taken from Pickles \&
 van der Kruit (1991) and supplementarily from Stanford, Eisenhardt, \&
 Dickinson (1995), although the latter gives only cluster memberships
 to which we assigned the cluster mean redshift $z=0.374$.
 To separate the sample of E/S0 galaxies, we use galaxy morphology as
 given by HST images (Stanford et al.\ 1995).
 As Abell~370 has a redshift $z=0.374$, the filter combination ($BVRI$) is
 expected to work to pick out cluster members to some extent as they
 bracket the 4000~{\AA} break (\S~2.2),
 although $U$-band is missing which is important to discriminate
 galaxies with lower B/T ratios at the cluster redshift from those with higher
 B/T ratio at lower redshifts.
 However, unfortunately, the photometric accuracy is poor, 
 especially in $B$ and $I$ bands
 (0.13$-$0.19 mag in $B$, 0.04$-$0.08 mag in $V$, 0.03$-$0.07 mag in $R$ and
 0.08$-$0.16 in $I$).

 We estimated the photometric redshifts for our sample galaxies.
 The results for field galaxies and cluster members are shown separately
 in Figs.~\ref{fig:a370_field} and~\ref{fig:a370_cluster}, respectively.
 The cluster members are defined as those which have spectroscopic redshifts
 $0.374-0.02<z<0.374+0.02$.
 As for the E/S0 galaxies, we can estimate the redshifts very well within
 $|\Delta z|<0.1$ both for field galaxies and cluster members.
 On the other hand, some disk galaxies have over- or under-estimated
 redshifts. In Fig.~\ref{fig:a370_field}, there is a galaxy 
 with $\Delta z>0.4$.
 That is because the photometry in $I$-band
 of this galaxy is very poor as indicated in the original table in
 Pickles \& van der Kruit (1991), and in fact if we use only $BVR$ bands for
 this galaxy, the estimated redshift agrees with the real redshift at the 1.5
 $\sigma$ level.
 The other two field disk galaxies are slightly underestimated by
 $\Delta z \sim -0.1$.
 These galaxies have very blue colours in $B-V$ or $V-R$, and it is suggested
 that they are either disk dominated galaxies or the ones strongly influenced
 by a recent star burst.
 For the cluster members (Fig.~\ref{fig:a370_cluster}) also, there are some
 galaxies whose redshifts are underestimated as much as $\Delta z \sim -0.15$.
 These galaxies tend to have low estimated B/T ratios and are again
 degenerate with galaxies at lower redshift.
 The discrimination between a blue cluster member and a slightly
 redder galaxy at lower redshift can be difficult, especially when we lack
 a bluer band corresponding to far-UV region (2500-3000~{\AA}).

\begin{figure}
\begin{center}
  \leavevmode
  \epsfxsize 1.0\hsize
  \epsffile{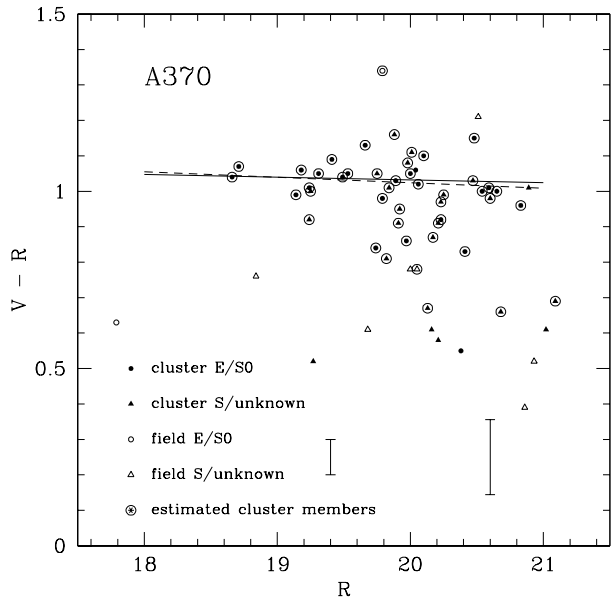}
\end{center}
\caption
{
The {\it C-M} diagram for Abell 370.
The filled symbols indicate cluster members, while the open symbols
show field galaxies. The symbols surrounded by a large circle indicate
the estimated cluster members selected using $|\Delta z|<0.1$.
Two error bars at the lower part of the figure indicate the typical
one sigma observational errors.
The solid and dashed lines indicate the {\it C-M} relation of
the the real cluster E/S0's and that of the estimated cluster E/S0's,
respectively, calculated using bi-weight fitting to the data.
}
\label{fig:a370_cm}
\end{figure}

\begin{figure}
\begin{center}
  \leavevmode
  \epsfxsize 1.0\hsize
  \epsffile{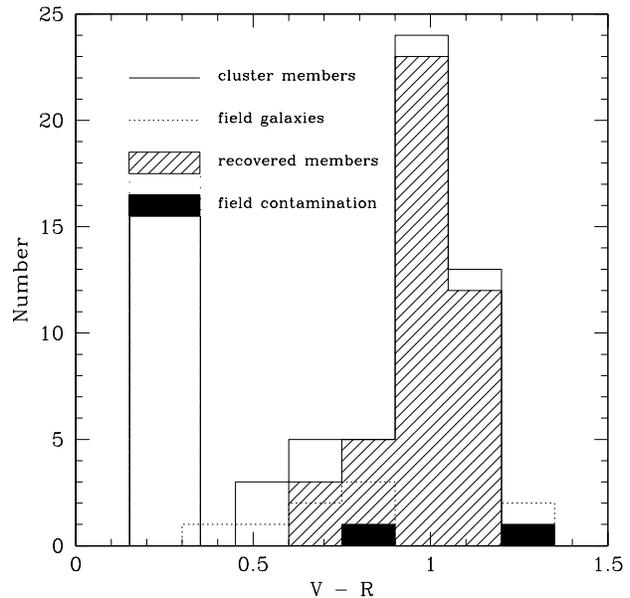}
\end{center}
\caption
{
The colour histogram for Abell 370.
The solid line shows the real cluster members, and the dotted line
shows the real field galaxies. The slantwise hatched area indicates
recovered cluster members, while the black shaded area indicates field
contamination.
It is shown that most of the cluster members are recovered, although a small
number of the bluest galaxies are dropped out. This would be 
improved if the $U$-band data were available.
Field contamination is also negligibly small.
}
\label{fig:a370_hist}
\end{figure}

 Nevertheless, if we adopt the criteria of defining cluster members
 as $|\Delta z| < 0.1$, we can pick out most of the cluster members
 with little field contamination as shown in Figs.~\ref{fig:a370_cm} and
 \ref{fig:a370_hist}. This is especially true for early-type galaxies.
 As a result, the {\it C-M} relation of the E/S0 galaxies is well recovered
 (Fig.~\ref{fig:a370_cm}). The solid line shows the real relation for the
 real cluster members while the dashed line shows the estimated relation
 for the estimated cluster members. We used a bi-weight fitting method
 to calculate these {\it C-M} relations (Beers, Flynn, \& Gebhardt 1990).
 Both relations are nearly identical.

 In summary, although the photometric accuracy is not ideal,
 we can still pick out most of the cluster members in A370, especially
 early-type galaxies, only photometrically based on our method.
 The field contamination is negligibly small.
 The method has difficulty in 
 recovering cluster members bluer than $V-R=0.6$, but this would
 be improved if $U$-band data were available.

\subsection{Hubble Deep Field}

\begin{figure*}
\begin{center}
  \leavevmode
  \epsfxsize 1.0\hsize
  \epsffile{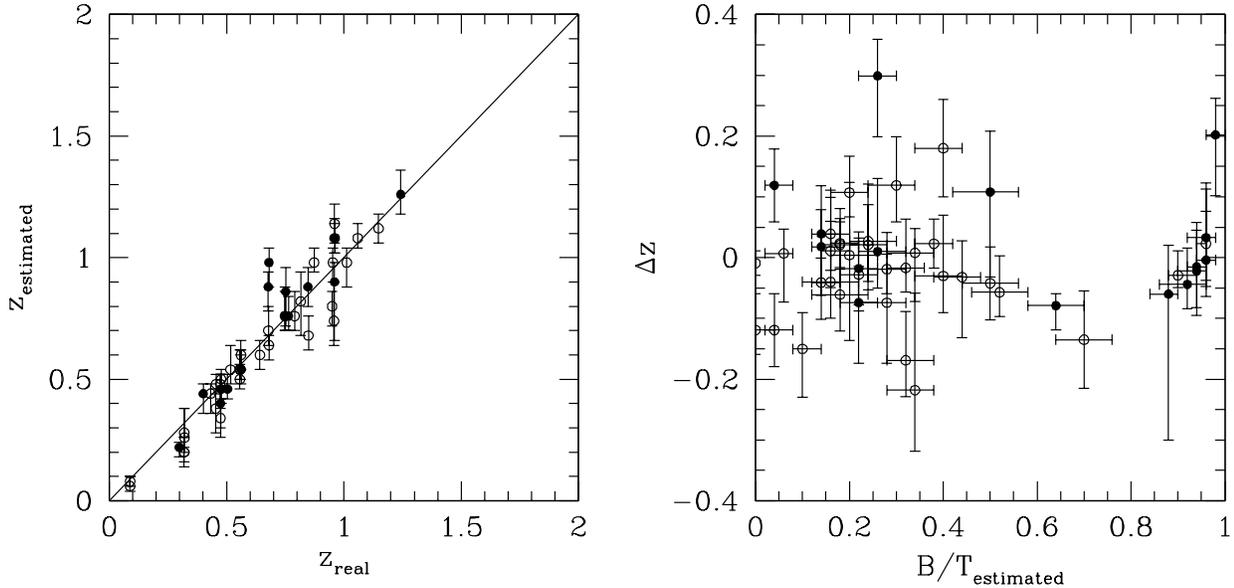}
\end{center}
\vspace{-9cm}
\caption
{HDF galaxies. $U300, B450, V606, I814$, and $J$ passbands are used.
Filled symbols indicate morphologically classified E/S0 galaxies,
while the rest show other type galaxies or unclassified galaxies.
}
\label{fig:hdf}
\end{figure*}

To further test our method, we apply it to galaxies taken from the
Hubble Deep Field.
The galaxies used here are chosen from Cowie's $K$-selected galaxy sample
(http://www.ifa.hawaii.edu/$^{\sim}$cowie/k\_table.html), all of which have
spectroscopic redshifts, mainly from Cohen et al.\ (1996).
Isophotal magnitudes in four HST WFPC2 filters (F300W[$U300$], F450W[$B450$],
F606W[$V606$], and F814W[$I814$])
are taken from Williams et al.\ (1996).
We cross-identify galaxies between Cowie's catalogue and the photometry
catalogue using RA and DEC. We choose only isolated galaxies to avoid
mis-identification. Also we excluded galaxies with $z>2$,
as the current passbands no longer bracket the 4000~{\AA} break.
Photometric errors are calculated from S/N ratios, but for those less
than 0.05 magnitude, we assume a minimum error of 0.05 magnitude.
We give larger minimum error of 0.2 magnitude to $U_{300}$ for galaxies with
$z_{\rm estimated}^{\rm best} > 0.2$, and to $B_{450}$ for galaxies with
$z_{\rm estimated}^{\rm best} > 0.8$,
in order to avoid incorporating uncertain model far-UV colours.
This is an iterative approach, but is unavoidable given the uncertainty of
the model UV spectrum.
The infrared photometry data in $J$ are acquired from Cowie's table.
A photometric accuracy of 0.1 magnitude is assumed 
in $J$-band as no information is given.

Applying the model directly to this data set, we find a systematic offset 
of $\Delta z \sim -0.1$ between the real redshifts and the estimated redshifts.
The most likely cause of this discrepancy is
a zero-point mismatch of order 0.1 magnitude between the data and the model
in such a way that the model is slightly redder in optical colours and
a bit bluer in far-UV colour.
It might be the intrinsic zero-point uncertainty in the model, since it is
comparable to the limitation of the current population synthesis models
(Charlot, Worthey, \& Bressan, 1996),
although it is puzzling that a similar problem is not seen in Abell~370.
To correct this situation, we shift the model zero-points for this case
only: \ie $+$0.1 magnitude is added onto the model $U300$, $I814$, and $J$.
With this zero-point shift, most of the redshifts of HDF galaxies are
correctly estimated as shown in Fig~\ref{fig:hdf}.
The RMS errors of the estimated redshift are smaller than 0.1;
\ie $\sigma(\Delta z)=0.091$ and $\sigma(error)=0.075$.
Although the zero-point mismatch is a problem, it is encouraging that our
method can estimate redshifts correctly over a wide range of redshifts.
However this exercise makes it
clear that the model should be calibrated with real data rather
than being applied blind to high redshift systems. This can be 
acheived using a handfull of spectroscopically confirmed members
of the target cluster without compromising the overall aim of
examining the star formation histories of the galaxy populations.

\section{Application to high redshift clusters}

\begin{table*}
\caption{Simulated galaxies in a $z=1$ cluster field.}
\label{tab:z1}
\begin{center}
\begin{tabular}{c|lcccc}
\hline\hline
& type & n & $\langle$[M/H]$\rangle_{\rm bulge}$ & $z_f$ & B/T \\
\hline
cluster & E/S0 & 10 & $0.06 \sim -0.52$ & 4.5            & 1.0 \\
        &      & 10 & $0.06 \sim -0.52$ & $4.5 \sim 1.5$ & 1.0 \\
        &      & 10 & $0.06 \sim -0.52$ & 4.5            & $1.0 \sim 0.5$ \\
        & Sp   & 20 & $0.06 \sim -0.52$ & 4.5            & $0.5 \sim 0.0$ \\
\hline
field & E            &  6 & $0.06 \sim -0.52$ & 4.5 & $1.0  \sim 0.6$ \\
      & S0           & 26 & $0.06 \sim -0.52$ & 4.5 & $0.6  \sim 0.5$ \\
      & Sab & 17 & $0.06 \sim -0.52$ & 4.5 & $0.5  \sim 0.15$ \\
      & Sc--Im  & 5 & $0.06 \sim -0.52$ & 4.5 & $0.15 \sim 0.05$ \\
\hline\hline
\end{tabular}
\end{center}
\end{table*}

We are interested in applying this classifier to high redshift clusters
around $z \gsim 1.0$, but, at present, a suitable data set is not avilable.
To show the applicability of the classifier to targets at that redshift,
we simulated a $z=1$ cluster field using the model described in \S~2.
Although this is a self-consistency
check (most importantly, it assumes that the photometric properties
of real galaxies are accurately described by the stellar population
synthesis code) it allows us to estimate the biases present in the
recovered galaxies samples and to determine how much
photometric accuracy and which combination of passbands is required to
pick out cluster members effectively in such a cluster.

We generated field galaxies using the type dependent prior 
distribution outlined in \S~3.2.
The metallicity of the bulge was chosen so that
$M_V^{\rm bulge}=-23, -20$ and $-17$ measured at $z=0$ corresponding to
$\langle$[M/H]$\rangle_{\rm bulge}=0.06, -0.23$, and $-0.52$.
Here we have simulated a $K$-limited galaxy sample
with $m_K<20$ for 1~arcmin$^2$ field of view which corresponds to 0.5~Mpc
$\times$ 0.5~Mpc at $z=1$, using the prior distribution, taking the
type-dependent LF into account.
The number of galaxies in each type is summarised in the lower half of
Table~\ref{tab:z1}.
As for the cluster members, we assumed the mix of galaxy populations given
in upper half of Table~\ref{tab:z1}; \ie 10 E/S0's from a metallicity sequence,
another 10 E/S0's from an age sequence, and the other 10 E/S0's from a B/T
sequence, and finally 20 disk galaxies are added in.
Firstly, by using $K$-band luminosity functions of high redshift cluster
galaxies (mean redshift 0.43) which are given separately for E/S0's and
Spirals (Barger et al.\ 1998), we assigned $K$-band absolute magnitude at
$z=0.43$ for a given galaxy.
Secondly, a formation epoch ($z_f$) and a B/T ratio are randomly assigned
in the respective range given in Table~\ref{tab:z1}.
Then we can assign its bulge metallicity using its $M_V^{\rm bulge}$ at $z=0$
calculated from $M_K$ at $z=0.43$.
If a galaxy has $m_K>20$ at $z=1$, it is rejected from our sample,
and the process is repeated until we finally obtain 50 cluster galaxies in
total.
We assigned the model magnitudes in various bands for each galaxy both
in the field and in the cluster. A Gaussian photometric 
error with $\sigma=0.071$ is
added on each colour of each galaxy.
We then regard these generated photometric data as the observational ones
for the $z=1$ cluster field, and the redshift classifier is applied to each
galaxy to estimate a redshift.

\subsection{Biases in the recovered galaxy properties}

\begin{figure*}
\begin{center}
  \leavevmode
  \epsfxsize 1.0\hsize
  \epsffile{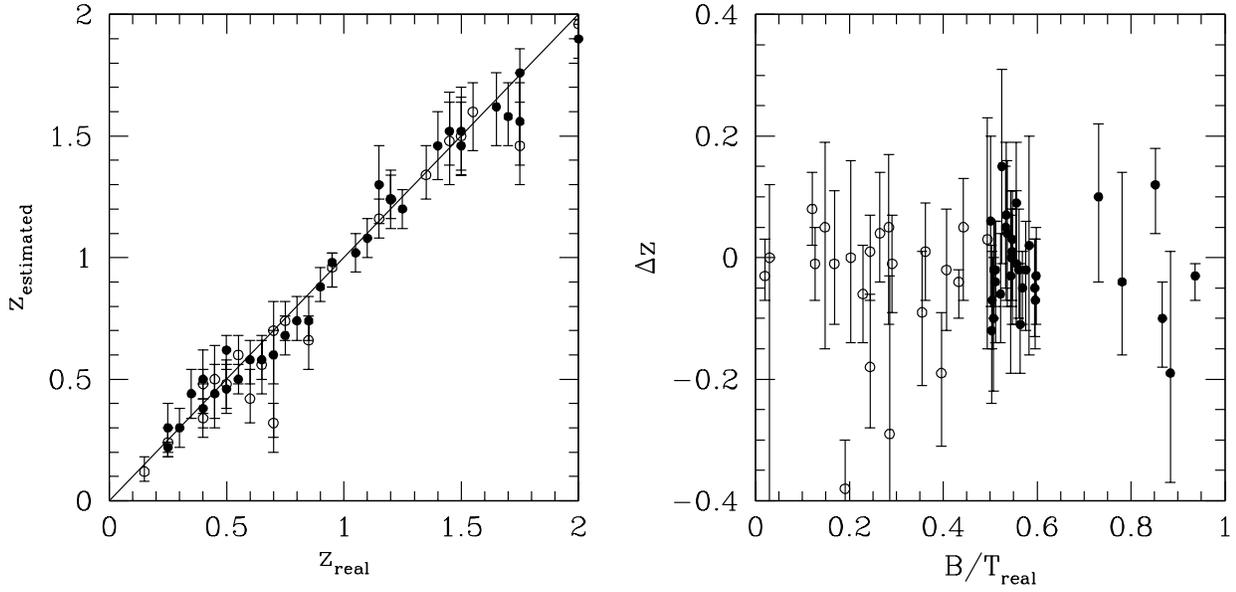}
\end{center}
\vspace{-9cm}
\caption
{
Field galaxies in the simulated cluster field at $z=1$.
Plotted are the estimated redshift vs. input real redshift (left), and
redshift error vs. input real B/T ratio (right).
Redshifts are estimated using $VRIK$ passbands with random Gaussian photometric
errors of $\sigma=0.071$ magnitude in all bands. 
Filled circles indicate E/S0 galaxies defined as B/T$\geq 0.5$,
while open circles indicate disk galaxies with B/T$<0.5$.
}
\label{fig:z1_field}
\end{figure*}

\begin{figure*}
\begin{center}
  \leavevmode
  \epsfxsize 1.0\hsize
  \epsffile{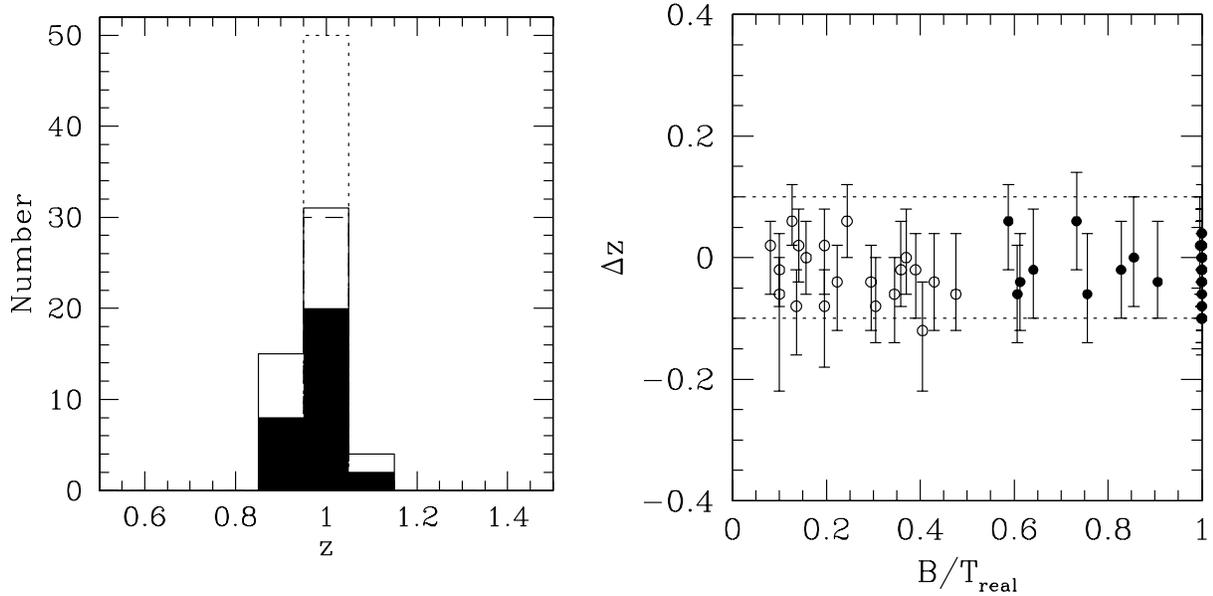}
\end{center}
\vspace{-9cm}
\caption
{
Cluster members in the simulated cluster field at $z=1$.
Distribution of the estimated redshifts (left) and redshift error vs.
input real B/T ratio (right).
Caption for the lines and the symbols are the same as
Fig.~\ref{fig:a370_cluster}.
}
\label{fig:z1_cluster}
\end{figure*}

\begin{figure}
\begin{center}
  \leavevmode
  \epsfxsize 1.0\hsize
  \epsffile{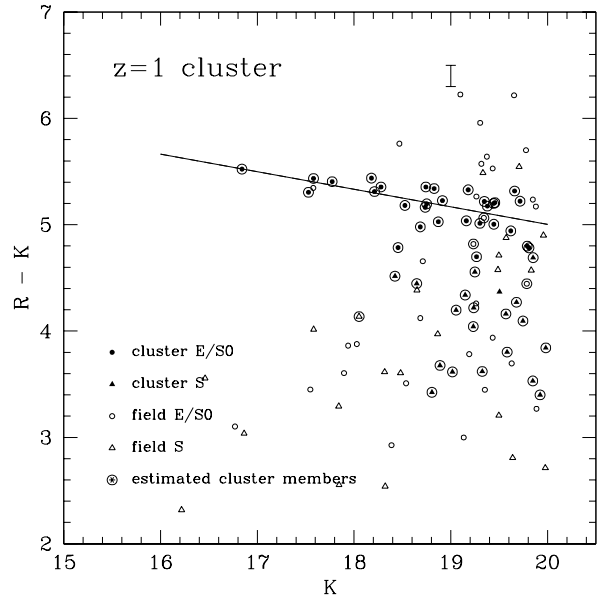}
\end{center}
\caption
{
The {\it C-M} diagram for a simulated cluster field at $z=1$.
Caption is the same as Fig.~\ref{fig:a370_cm}.
Note that the real {\it C-M} relation and the estimated one are identical
(solid line).
}
\label{fig:z1_cm}
\end{figure}

\begin{figure}
\begin{center}
  \leavevmode
  \epsfxsize 1.0\hsize
  \epsffile{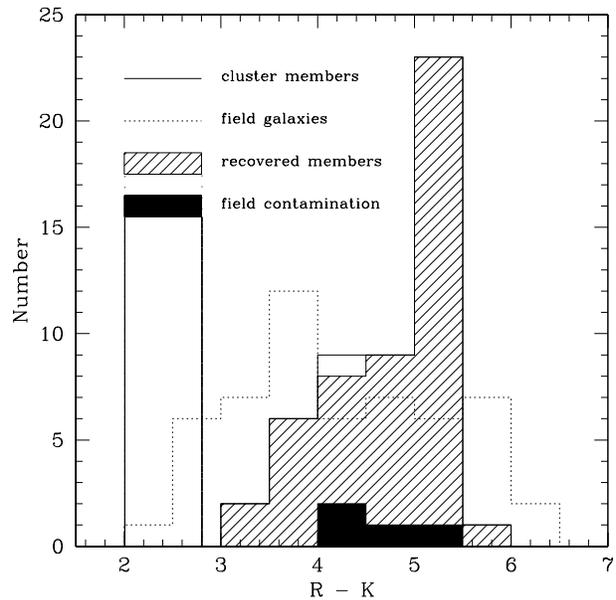}
\end{center}
\caption
{
The colour histogram of a simulated cluster field at $z=1$.
The solid line shows the real cluster members, and the dotted line
shows the real field galaxies. The slantwise hatched area indicates
recovered cluster members, while the black shaded area indicates field
contamination.
There are few dropped out members and little field contamination,
and importantly they have no bias in colours.
}
\label{fig:z1_hist}
\end{figure}

The results for the field galaxies and the cluster members are shown
separately in Figs.~\ref{fig:z1_field} and \ref{fig:z1_cluster}.
We used $VRIK$ colours to estimate the redshifts. $B$-band was not used, since
at $z=1$ it falls in the far-UV spectral region, well below 2500~{\AA}.
The overall agreement between the estimated redshift and the real redshift
is excellent. Most of the galaxy redshifts are well recovered within
$|\Delta z|=0.1$, regardless of real redshift and irrespective of galaxy type.
As a result, as shown in Fig~\ref{fig:z1_cm}, 
the recovery of cluster members is magnificent.
Here we adopted the criterion of cluster members as $|\Delta z|<0.1$,
considering the photometric accuracy (see Fig.\ref{fig:error}).
It is also adequate since it does not pick up large amount of field
contamination at high redshifts.
In this case, only one cluster spiral is dropped out.
Field contamination is also negligible (only four galaxies).
We have recovered not only old ellipticals, but also young or star forming
ellipticals and spirals as well.
To see the bias in the identification of cluster 
members as a function of galaxy colour,
we show the colour histogram of the recovered cluster members and the field
contamination in Fig.~\ref{fig:z1_hist}. As is clear from the figure,
there is no colour bias at all in either the cluster or the field. 
Consequently, we recover the {\it C-M} relation of E/S0 galaxies very well
(identically in this case), as shown by the solid line in the figure.
The bi-weight scatters around the relation are also calculated and given in
Table~\ref{tab:scatter} as well as the values of the {\it C-M} slope.
The numbers for Abell~370 are also given in the same table.
Both scatters and slopes are almost correctly estimated irrespective of
galaxy type.

\begin{table}
  \caption{The bi-weight scatters and slopes of the {\it C-M} relation.
the values for all types and E/S0 galaxies are shown separately.
The scatters are measured with respect to the {\it C-M} relation of E/S0
galaxies only.
}
  \label{tab:scatter}
  \begin{tabular}{lccccc}
   \hline
   & & \multicolumn{2}{c}{Abell~370} & \multicolumn{2}{c}{$z=1$ cluster}\\
   & & all & E/S0 & all & E/S0 \\
   \hline
   scatter & real      & 0.125 & 0.073 & 0.613 & 0.169 \\
           & estimated & 0.112 & 0.072 & 0.603 & 0.189 \\
   \hline
   slope   & real      & $-$0.018 & $-$0.008 & $-$0.423 & $-$0.165 \\
           & estimated & $-$0.018 & $-$0.016 & $-$0.381 & $-$0.165 \\
   \hline
  \end{tabular}
  \medskip
\end{table}

All the above results are encouraging.
If the models were perfect, we could assign redshift with
0.1 accuracy or better with $<$0.1 mag photometric errors in all bands.
With this success, we will be able to extend the {\it C-M} relation analysis
(eg., Kodama et al.\ 1998; Ellis et al.\ 1997; Stanford, Eisenhardt, \&
Dickinson 1998)
to $z\gsim1$ clusters without taking spectroscopic redshifts.
Importantly, we can pick out cluster galaxies with various stellar
populations; \ie not only passively evolving old galaxies but also the
galaxies which have a significant contribution from younger 
stellar populations.
This is encouraging, as it is important to select the cluster members with as
little bias as possible.
Our method will allow us to conduct the colour scatter analysis around
the {\it C-M} relation reliably, and to look into the age dispersion of cluster
galaxies at high redshifts, if any.

\subsection{Optimal passbands for cluster identification}

\begin{table*}
  \caption{Passband choice for a simulated cluster field at $z=1$. 
   Numbers of dropped out members and field contamination are also presented.
   Percentage of dropped members and field contamination are defined per real
   cluster members and per estimated cluster members, respectively.}
  \label{tab:band2}
  \begin{tabular}{ccccccccc}
    \hline
  & \multicolumn{2}{c}{all} & \multicolumn{2}{c}{E/S0} & \multicolumn{2}{c}{dropped members} & \multicolumn{2}{c}{field contamination}\\
  passbands & $\sigma(\Delta z)$ & $\sigma(error)$ & $\sigma(\Delta z)$ & $\sigma(error)$ & all & E/S0 & all & E/S0\\
  \hline
  $BVRIK$ & 0.070 & 0.074 & 0.068 & 0.074 & 3 (6\%) & 3 (10\%) & 5 (10\%) & 4 (13\%)\\
  $VRIK$  & 0.079 & 0.099 & 0.064 & 0.095 & 1 (2) & 0 (0) & 4 (8) & 3 (9)\\
  $VIK$   & 0.141 & 0.129 & 0.069 & 0.112 & 7 (14) & 1 (3) & 4 (9) & 3 (9)\\
  $VRK$   & 0.114 & 0.131 & 0.087 & 0.129 & 4 (8) & 2 (7) & 3 (6) & 3 (10)\\
  $VRI$   & 0.231 & 0.232 & 0.243 & 0.216 & 9 (18) & 3 (10) & 10 (20) & 7 (21)\\
  $RIK$   & 0.155 & 0.133 & 0.074 & 0.118 & 16 (32) & 3 (10) & 3 (8) & 3 (10)\\
  $RJK$   & 0.211 & 0.212 & 0.159 & 0.216 & 38 (76) & 21 (70) & 7 (37) & 4 (31)\\
  \hline
  \end{tabular}
  \medskip
\end{table*}

In Table~\ref{tab:band2}, we summarise the effect of passband choice on the
estimated redshift error.  Photometric errors of 0.071 mag are
assumed in all passbands.
Two cases in particular are found to be ideal: $BVRIK$ and $VRIK$.  The
redshift errors are always well below 0.1 regardless of galaxy type,
and hence the number of galaxies dropped out of our cluster sample, and
the field contamination, are minimised.  It is encouraging that we can
do comparably well without $B$-band, as it is good to minimise the use
of passbands shortwards of rest frame 2500~{\AA} where possible.  It
should be noted that $VIK$ and $VRK$ work comparably well, although 
$\sigma(error)$ is larger, especially for galaxies with ongoing star
formation.  This is because it becomes difficult to separate the
effects of changes in stellar population and those of changing
redshift.  If both $B$ and $V$-band are missing (and we have $RIK$
only), we tend to underestimate the redshift of bluer galaxies.
This is analogous to lacking $U$-band for Abell 370: it becomes
difficult to disentangle galaxy type and redshift without the UV
colours for star forming galaxies.  For the early type galaxies,
however, the redshift errors are reasonable, as we would expect from
Fig. \ref{fig:twocolour}.  In contrast, if $K$ band is missing, the
errors in the redshift estimation become much larger, irrespective of
galaxy type, as there is no passband longwards of the rest-frame 4000~{\AA}
break.  In this context, it is crucial for high redshift work to
have both optical and near infrared passbands for accurate redshift
estimation.  

\section{Summary}

We present a new photometric redshift estimator, which is optimised
for the identification and study of galaxy clusters at high redshifts.
We use only several broad passbands covering the 4000~{\AA} break, and
find in practice that it is possible to avoid the use of the uncertain
colours shortwards of rest frame 2500~{\AA}.
In our models, we considered as 
wide a variety of stellar populations as is possible
to minimise the selection bias in the recovered
cluster members.
As most of the effects of changing stellar population on the integrated
colours are highly degenerate, we find that it is possible to 
estimate redshifts with reasonable
accuracy for a range of galaxy types, ranging from those with old,
passively evolving stellar populations, through to those with younger
stellar populations and on-going star formation.

Following the success in testing our method with data from Abell~370 and
from the Hubble Deep Field, we applied it to a simulated cluster at $z=1$.
We have shown that the estimation of redshifts with accuracies better
than  $|\Delta z|<0.1$
can be achieved with multi-passband photometry of moderate quality
($\lsim$ 0.1 mag) in a small number of passbands, and the cluster members
can be reliably identified. Therefore, the recovery of the {\it C-M} relation
both in terms of the slope and scatter is expected to be 
accurate and almost free from any selection bias.
We now have a means of analysing the {\it photometric} properties of cluster
galaxies at very high redshifts without a thorough spectroscopic membership
confirmation.

\section*{Acknowledgements}

We would like to thank C. Tadhunter and S. Warren for useful discussion.
We are also grateful to the anonymous referee who gave us many constructive
comments.
T.K. thanks JSPS Postdoctoral Fellowships for Research Abroad for
financial support.  E.F.B. would like to thank the Isle of Man
Education Department for their generous financial support.
This project made use of STARLINK computing facilities at Durham and
Cambridge.


\begin{thebibliography}{}
 
\bibitem[]{} Allington-Smith, J., et al.,
     1994, PASP, 106, 983
\bibitem[]{} Arimoto, N., 1996, in: From Stars to Galaxies, eds. C. Leitherer,
     U. Fritze-von Alvensleben, J. Huchra (ASP Conf. Ser. Vol. 98), p.287
\bibitem[]{} Barger, A. J., Arag\'on-Salamanca, A., Smail, I., Ellis, R. S.,
     Couch, W. J., Dressler, A., Oemler, A., \& Poggianti, B. M.,
     1998, ApJ, in press
\bibitem[]{} Baugh, C. M., Cole, S.,  Frenk, C. S., \& Lacey, C. G,
     1998, ApJ, in press
\bibitem[]{} Beers, T. C., Flynn, K., \& Gebhardt, K., 1990, AJ, 100, 32
\bibitem[]{} Berlind, A. A., Quillen, A. C., Pogge, R. W., \& Sellgren K.,
     1997, AJ, 114, 107
\bibitem[]{} Bica, E., 1988, A\&A, 195, 76
\bibitem[]{} Bingelli, B., Sandage, A., \& Tammann, G. A., 1988,
     ARA\&A, 26, 509
\bibitem[]{} Bower, R. G., Lucey, J. R., \& Ellis, R. S., 1992a, MNRAS, 254,
     589
\bibitem[]{} Bower, R. G., Lucey, J. R., \& Ellis, R. S., 1992b, MNRAS, 254,
     601
\bibitem[]{} Bower, R. G. \& Smail, I., 1997, MNRAS, 290, 292
\bibitem[]{} Bromley, B. C. , Press, W. H., Lin, H., \& Kirschner, R. P.,
     1997, astro-ph/9711227
\bibitem[]{} Bruzual, G. A. \& Charlot, S. 1993, ApJ, 405, 538
\bibitem[]{} Burstein, D., Bertola, F., Buson, L. M., Faber, S. M.,
     \& Laner, T., 1988, ApJ, 328, 440
\bibitem[]{} Buta, R., Mitra, S., de Vaucouleurs, G., \& Corwin, Jr. H. G.,
     1994, AJ, 107, 118
\bibitem[]{} Buta, R. \& Williams, K. L., 1995, AJ, 109, 543
\bibitem[]{} Calzetti, D., Kinney, A. L., \& Storchi-Bergmann, T., 1994,
     ApJ, 429, 582
\bibitem[]{} Carroll, S. M., Press, W. H., Turner, E. L., 1992, ARA\&A, 30, 499
\bibitem[]{} Charlot, S., Worthey, G., \& Bressan, A., 1996, ApJ, 457, 625
\bibitem[]{} Cohen, J. G., Cowie, L. L., Hogg, D. W., Songaila, A.,
         Blandford, R., Hu, E. M., Shopbell, P., 1996, ApJ, 471, 5
\bibitem[]{} Connolly, A. J., Csabai, I., Szalay, A. S., Koo, D. C.,
     Kron, R. G., \& Munn, J. A., 1995, AJ, 110, 2655
\bibitem[]{} Cowie, L. L., Songaila, A., Hu, E. M., \& Cohen, J. G., 1996,
     AJ, 112, 839
\bibitem[]{} de Jong, R. S., 1996, A\&A, 313, 377
\bibitem[]{} de Vaucouleurs, G., de Vaucouleurs, A., Corwin, H. G.,
     Buta, R. J., Paturel, G., \& Fouqu\'e, P., 1991, Third Reference
     Catalog of Bright Galaxies (Springer;Berlin) (RC3)
\bibitem[]{} Deltorn, J.-M., Le Fevre, O., Crampton, D., \& Dickinson, M.,
     1997, ApJ, 483, 21
\bibitem[]{} Ellis, R. S., Smail, I., Dressler, A., Couch, W. C.,
     Oemler Jr, A., Butcher, H., \& Sharples, R. M., 1997,
     ApJ, 483, 582
\bibitem[]{} Glazebrook, K., Ellis, R. S., Colles, M., Broadhurst, T.,
     Allington-Smith, J. J., \& Tanvir, N., 1995, MNRAS, 273, 157
\bibitem[]{} Gwyn, S. D. J. \& Hartwick, F. D. A., 1996, ApJ, 468, L77
\bibitem[]{} Hammer, F., Flores, H., Lilly, S. J., Crampton, D.,
     Le F\`evre, O., Rola, C., Mallen-Ornelas, G., Schade, D., \& Tresse, L.,
     1997, ApJ, 481, 49
\bibitem[]{} Jablonka, P., Martin, P., \& Arimoto, N., 1996, AJ, 112, 1415
\bibitem[]{} Jones, H. \& Bland-Hawthorn, J., 1997, PASA, 14, 8
\bibitem[]{} Kennicutt, R. C., 1992, ApJS, 79, 255
\bibitem[]{} Kodama, T. \& Arimoto N., 1997, A\&A, 320, 41
\bibitem[]{} Kodama, T., 1997, Ph.D. Thesis, Univ. of Tokyo
\bibitem[]{} Kodama, T., Arimoto, N., Barger, A. J., \& Arag\'on-Salamanca, A.,
     1998, A\&A, 334, 99
\bibitem[]{} K\"oppen, J. \& Arimoto, N., 1990, A\&AS, 240, 22
\bibitem[]{} Lanzetta, K. M., Yahil, A., \& Fern\'{a}ndez-Soto, A., 1996,
     Nature, 381, 759
\bibitem[]{} Madau, P., Ferguson, H. C., Dickinson, M. E., Giavalisco, M.,
     Steidel, C. C., \& Fruchter, A., 1996, MNRAS, 283, 1388
\bibitem[]{} Marzke, R. O., Geller, M. J., Huchra, J. P., \& Corwin Jr.,
     H. G., 1994, AJ, 108, 437
\bibitem[]{} Marzke, R. O., da Costa, L. N., Pellegrini, P. S.,
     Willmer, N. A., \& Geller, M. J., 1998, astro-ph/9805218
\bibitem[]{} Mathis, J. S., 1990, ARA\&A, 28, 37
\bibitem[]{} Metcalfe, N, Shanks, T., Campos, A.,  Fong, R., \& Gardner, J.P., 
     1996. Nature, 383, 236 
\bibitem[]{} Mobasher, B., Rowan-Robinson, M., Georgakakis, A., \& Eaton, N. 
     1996, MNRAS, 282, 7
\bibitem[]{} Nomoto, K., 1993, private communication
\bibitem[]{} Pickles, A. J. \& van der Kruit, P. C., 1991, A\&AS, 91, 1
\bibitem[]{} Salpeter, E. E., 1955, ApJ, 121, 161
\bibitem[]{} Sawicki, M. J., Lin, H., \& Yee, H. K. C., 1997, AJ, 113, 1
\bibitem[]{} Schechter, P. L., 1976, ApJ, 203, 297
\bibitem[]{} Schmidt, M., 1959, ApJ, 129, 243
\bibitem[]{} Shimasaku, K. \& Fukugita, M., 1998, ApJ, in press
\bibitem[]{} Simien, F. \& de Vaucouleurs, G., 1986, ApJ, 302, 564
\bibitem[]{} Stanford, S. A., Eisenhardt, P. R. M., \& Dickinson, M., 1995,
     ApJ, 450, 512
\bibitem[]{} Stanford, S. A., Eisenhardt, P. R. M., \& Dickinson, M., 1998,
     ApJ, 492, 461
\bibitem[]{} Stanford, S. A., Elston, R., Eisenhardt, P. R., Spinrad, H.,
     Stern, D., \& Dey, A., 1997, AJ, 114, 2232
\bibitem[]{} Steidel, C. C., Adelberger, K. L., Dickinson, M.,
     Giavalisco, M., Pettini, M., \& Kellogg, M., 1998, ApJ, 492, 428
\bibitem[]{} Steidel, C. C., Giavalisco, M., Pettini, M., Dickinson, M.,
     \& Adelberger, K. L., 1996, ApJ, 462, 17
\bibitem[]{} Taylor, K., 1995, A\&AS, 186, 4406
\bibitem[]{} Tinsley, B. M., 1980, Fundamentals of Cosmic Physics, Vol.5, p.287
\bibitem[]{} White III R.E., Keel W.C., \& Conselice C.J., 1996, 
     astro-ph/9608113  
\bibitem[]{} Williams, R. E., et al., 1996, AJ, 112, 1335
\bibitem[]{} Worthey, G., 1994, ApJS, 95, 107
\bibitem[]{} Yamada, T., Tanaka, I., Arag\'on-Salamanca, A., Kodama, T.,
     Ohta, K., \& Arimoto, N., 1997, ApJ, 487, L125
\bibitem[]{} Yi, S., Demarque, P., \& Oemler, Jr. A., 1997, ApJ, 486, 201
\bibitem[]{} Young, J. S. \& Knezek, P. M., 1989, ApJ, 347, L55

\end{thebibliography}
\end{document}